\documentclass[pra,onecolumn,superscriptaddress,floatfix,amssymb,showpacs]{revtex4-1}
\usepackage{graphicx}
\usepackage{bm}
\usepackage[tbtags]{amsmath}
\usepackage{verbatim}
\usepackage{dsfont}
\usepackage{cancel}
\usepackage{enumerate}
\usepackage{color}
\usepackage{mathrsfs}
\usepackage{braket}
\usepackage{bbold}

\newcommand{\tabincell}[2]{\begin{tabular}{@{}#1@{}}#2\end{tabular}}
\newcommand\red[1]{{\color{red}{#1}}}
\newcommand\blue[1]{{\color{blue}{#1}}}

\begin{document}
\title{%
	Quantum walk on a chimera graph
	}
\author{Shu Xu}
\affiliation{%
	Shanghai Branch, National Laboratory for Physical Sciences at Microscale,
    	University of Science and Technology of China, Shanghai 201315, China%
	}
\author{Xiangxiang Sun}
\affiliation{%
	Department of Modern Physics,
	University of Science and Technology of China, Hefei, Anhui 230026, China%
    }
\author{Jizhou Wu}
\affiliation{%
    Shanghai Branch, National Laboratory for Physical Sciences at Microscale,
    University of Science and Technology of China, Shanghai 201315, China%
    }
\author{Wei-Wei Zhang}
\affiliation{%
    Shanghai Branch, National Laboratory for Physical Sciences at Microscale,
    University of Science and Technology of China, Shanghai 201315, China%
    }
\author{Nigum Arshed}
\affiliation{%
    Shanghai Branch, National Laboratory for Physical Sciences at Microscale,
    University of Science and Technology of China, Shanghai 201315, China%
    }
\author{Barry C. Sanders}
\email{bsanders@ustc.edu.cn}
\affiliation{%
    Shanghai Branch, National Laboratory for Physical Sciences at Microscale,
    University of Science and Technology of China, Shanghai 201315, China%
    }
\affiliation{%
    Synergetic Innovation Center
        of Quantum Information and Quantum Physics,
    University of Science and Technology of China, Hefei, Anhui 230026, China
    }
\affiliation{%
    Institute for Quantum Science and Technology, University of Calgary, Alberta, Canada T2N 1N4%
    }
\affiliation{%
    Program in Quantum Information Science,
    Canadian Institute for Advanced Research,
    Toronto, Ontario M5G 1Z8, Canada%
    }
\date{\today}
\begin{abstract}
We analyze a continuous-time quantum walk on a chimera graph,
which is a graph of choice for designing quantum annealers,
and we discover beautiful quantum walk features
such as localization that starkly distinguishes classical from quantum behaviour.
Motivated by technological thrusts,
we study continuous-time quantum walk on enhanced variants of the chimera graph
and on diminished chimera graph with a random removal of vertices.
We explain the quantum walk by constructing a generating set for a suitable subgroup of graph isomorphisms and corresponding symmetry operators that commute with the quantum walk Hamiltonian;
the Hamiltonian and these symmetry operators provide a complete set of labels for the spectrum and the stationary states.
Our quantum walk characterization of the chimera graph and its variants
yields valuable insights into graphs used for designing quantum-annealers.
\end{abstract}
\maketitle
\section{Introduction}
The chimera graph~\cite{boothby2016}, which we henceforth call~$\chi$,
is the hardware graph for D-Wave computers.
This graph
underpins the chip design for computational tasks such as
optimization~\cite{neukart2017,rosenberg2016,daryl2015},
graph partitioning~\cite{ushijima2017},
and machine learning~\cite{omalley2017,potok2016,adachi2015},
which classical computers and algorithms struggle to perform efficiently and accurately.
D-Wave computers solve problems by converting the problem graph into its hardware graph, 
using the minor embedding technique
of representing one logical qubit by several physical qubits~\cite{choi2008,choi2011,klymko2014}.
Long-range connections in~$\chi$~\cite{katzgraber2014,vinci2015}
are absent due to implementation constraints~\cite{bunyk2014},
which necessitates an embedding of problem graphs into the incomplete~$\chi$ graph
using the minor embedding technique~\cite{vinci2015,bunyk2014,venturelli2015}.
Therefore, modifying the current D-Wave computers graph structure is important,
 and studying the symmetries of~$\chi$ may offer another perspective to give insights to reduce overheads or improve minor embedding technology.

The quantum walk (QW)~\cite{aharonov1993},
which quantizes the ubiquitous classical random walk~\cite{noh2004},
has become a rich area of theoretical~\cite{childs2004,childs2009,lovett2010} and experimental~\cite{dur2002,zahringer2010,izaac2017} study
for quantum computation~\cite{farhi1998,daniel2010},
quantum search~\cite{ambainis2004},
photosynthetic energy transfer~\cite{masoud2008,biggerstaff2016,carrega2016},
topological phases~\cite{kitagawa2010,asboth2012},
quantum algorithms~\cite{kendon2006,santha2008},
quantum transport~\cite{mulken2005,bougroura2016}
and the foundational quantum-classical divide~\cite{childs2002}.
Two equivalent versions are the discrete
QW~\cite{aharonov2001,shikano2010}
and the continuous-time QW~\cite{farhi1998,mulken2011}.
For the discrete QW, the evolution operator is applied in discrete time steps,
whereas the evolution operator is defined for all times for the continuous-time QW~\cite{venegas-andraca2012}.
In both discrete and continuous models,
the topology on which QW is performed and its properties studied are discrete graphs~\cite{venegas-andraca2012}.
The topology of the graph has significant affect in the evolution.

Through simulating the continuous-time QW on~$\chi$ and its variants with different boundary conditions and initial positions,
we discover that QW localization behaviour is starkly different from the classical walk.
We study symmetries of~$\chi$ and the role they play in state evolution.
The properties of the walker's evolution on~$\chi$ reflect the structures of the graphs thereby indicating that our QW approach could be used for state engineering and quantum-gate design.
Our method provides a way to investigate the structure of graphs and
sheds light on how to understand the symmetries of~$\chi$,
which is used on the D-Wave machine.
We are studying a single-particle QW,
and quantum annealing on a~$\chi$ graph involves the ground-state of a many-body Hamiltonian so how much we can learn from the single-particle QW is inherently limited.
The nature of the many-body ground-state problem for classical Ising models~\cite{ceperley1980} and quantum model~\cite{white1992}
provides a rigorous and valuable complementary analyses for many-body cases.

The structure of our article is as follows.
We provide relevant background regarding the continuous-time QW and~$\chi$ in Section~\ref{sec:background}.
In Section~\ref{sec:result} we present our numerical simulation of the continuous-time QW on~$\chi$ and its variants.
We discuss the graph symmetries and spectra of the QW Hamiltonian in Section~\ref{sec:symmetries}.
In Section~\ref{sec:discussion}, we discuss the QW behaviour on chimera graph and its variants.
We present our conclusions in Section~\ref{sec:conclusion}.

\section{Background}
\label{sec:background}
In this section, we give a brief introduction regarding the continuous-time QW and~$\chi$.

\subsection{Continuous-time quantum walk}
The QW was originally motivated by the widespread use of the classical random walk in the design of randomized algorithms~\cite{kendon2006,santha2008}.
A quantum walker's evolution on a graph is described by the Schr$\ddot{\text{o}}$dinger equation~\cite{venegas-andraca2012}.
In the continuous-time QW,
a single-particle evolves over a connected graph~$\textsf{G}=(\textsf{V},\textsf{E})$,
for~$\textsf{V}=\{\textsf{v}\}$ the set of vertices of the graph
and~$\textsf{E}=\{\textsf{e}\}$
the set of edges where each edge~$\textsf{e}$ can be expressed by the pair of vertices~$(\textsf{v},\textsf{v}')$ it connects.
The Hamiltonian~$\hat{H}$
determining the walker's evolution
corresponds to the adjacency matrix representation of the edges,
and the edges can have weights~$\{w(\textsf{e})\}$,
which are the couplings between pair of vertices.

The bridge from graphs to quantum dynamics
is achieved by constructing the orthogonal vertex basis~$\{\ket{\textsf{v}}\}$.
The continuous-time QW evolution is generated by the adjacency matrix,
serving as the Hamiltonian
\begin{equation}
\label{eq:walkhamiltonian}
	H_{\textsf{v}\textsf{v}'}
		:=\bra{\textsf{v}}\hat{H}\ket{\textsf{v}'}
		=\begin{cases}
			h_{\textsf{v}},&\textsf{v}=\textsf{v}', \\
			-j_{\textsf{v}\textsf{v}'},&(\textsf{v},\textsf{v}')\in\textsf{E}_\text{intra}, \\
			-k_{\textsf{v}\textsf{v}'},& (\textsf{v},\textsf{v}')\in\textsf{E}_\text{inter}, \\
			0,&\text{otherwise},\\
        \end{cases}
\end{equation}
with~$h_{\textsf{v}}$ the onsite frequency term,
$j_{\textsf{v}\textsf{v}'}$ and
$k_{\textsf{v}\textsf{v}'}$ the transition rates between vertices
(edge weights), and
$\textsf{E}_\text{intra}$ and $\textsf{E}_\text{inter}$ the intracell and intercell edges.
Later we allow~$h$,~$j$ and~$k$ to be time-dependent to study adiabatic evolution~\cite{farhi2000},
but initially we consider a time-independent~$\hat{H}$
whose rows and columns add to zero
(thereby constraining~$\{h_{\textsf{v}}\}$)
in order to generate an evolution operator whose rows and columns sum to one
analogous to the properties of a bistochastic matrix~\cite{marshall1979}.

For a time-independent ~$\hat{H}$,
the unitary evolution operator is~$U=\text{e}^{-\text{i}\hat{H}t}$
starting at time~$t=0$ ($\hbar\equiv1$),
and the evolving walker wavefunction is~$\ket{\Psi(t)}$.
If the walker starts at the vertex $\ket{\textsf{v}}$ then the walker's state at time~$t$ is described by
\begin{equation}
\label{eq:walkunitary}
	\ket{\Psi(t)}
		=U\ket{\textsf{v}}=\text{e}^{-\text{i}\hat{H}t}\ket{\textsf{v}}.
\end{equation}
Unlike the classical random walk~\cite{noh2004},
the walker in the QW is in a superposition of vertices.
The transition amplitude $\alpha_{\textsf{v}\textsf{v}'}(t)$
from the vertex $\ket{\textsf{v}}$ to $\ket{\textsf{v}'}$ at time~$t$ is
\begin{equation}
\label{eq:walktransitionamplitude}
	\alpha_{\textsf{v}\textsf{v}'}(t)
		=\left\langle\textsf{v}' \left| \text{e}^{-\text{i}\hat{H}t} \right| \textsf{v} \right\rangle,
\end{equation}
and the corresponding transition probability $P_{\textsf{v}\textsf{v}'}(t)$ is~\cite{mulken2005}
\begin{equation}
\label{eq:walktransitionprob}
P_{\textsf{v}\textsf{v}'}(t):={|\alpha_{\textsf{v}\textsf{v}'}|^{2}}.
\end{equation}
The long-time average of~$P_{\textsf{v}\textsf{v}'}(t)$,
which is known as the limiting probability~\cite{mulken2011}
\begin{equation}
\label{eq:limitingprob}
	\bar{P}_{\textsf{v}\textsf{v}'}
		:={\lim_{\tau\rightarrow{\infty}}}\frac{1}{\tau}\int_{0}^{\tau}\text{d}t~P_{\textsf{v}\textsf{v}'}(t),
\end{equation}
embodies a natural notion of QW convergence and captures the amount of time the walker spends in each subset of the nodes~\cite{aharonov2001}.

\subsection{Chimera Topology}
\begin{figure}
\includegraphics[width=0.6\columnwidth]{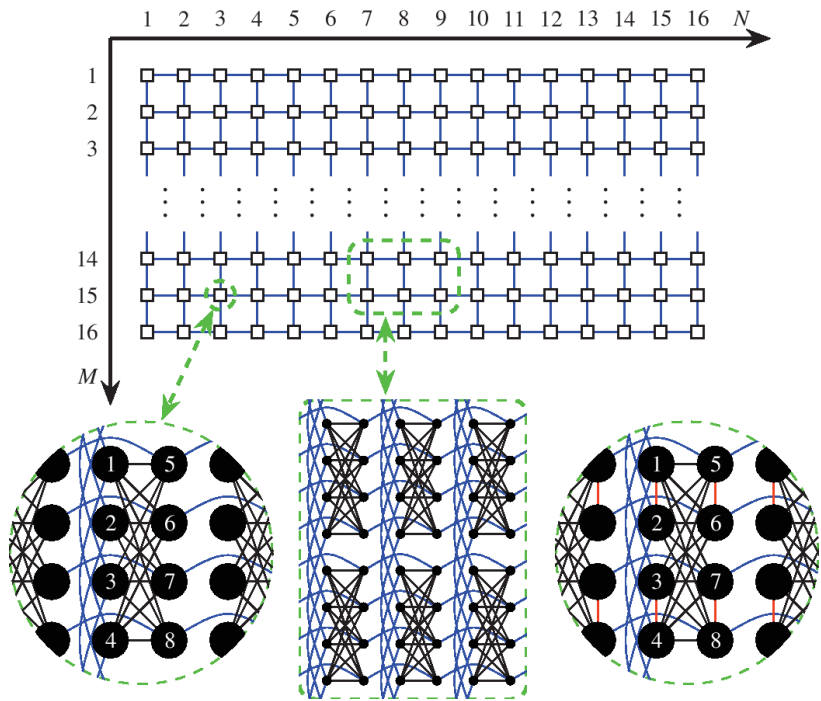}
\caption{%
        Chimera graph~$\chi_{16,16,4}$ with~$\square$ denoting unit cells.
       The left inset shows intracell coordinates and connectivity of~$\chi$.
       The middle inset shows intercell connectivity.
       The right inset shows intracell coordinates and connectivity of~$\chi^+$.%
       }
\label{fig:Chimeragrid16x16}
\end{figure}

The chimera graph
$\textsf{G}=\chi$,
is the underlying topology for the D-Wave Two,
D-Wave 2X,
and D-Wave 2000Q~\cite{king2017,boothby2016}.
D-Wave Systems'~$M \times N$ grid of unit cells is realized as a rectangular~$L\times2$ qubit array~\cite{boothby2016},
denoted here as~$\chi_{MNL}$.
The structure of~$\chi$ is shown
in Fig.~\ref{fig:Chimeragrid16x16},
which has~$MN$ unit cells and~$2MNL$ qubits.
The chimera vertices shown in Fig.~\ref{fig:Chimeragrid16x16} are noted as
\begin{equation}
\label{eq:VMNL}
	\textsf{V}_{MNL}=\{(m,n,\mu)\},\;
	m\in[1,M],
	n\in[1,N],\;
	\mu\in[1,2L],
\end{equation}
for~$m,n$ the cartesian coordinates of the unit cells
and~$\mu$
the vertex label for the~$L\times2$ cell
with~$\mu\leq L$ for left side
and~$\mu >L$ for the right side.

The edge set is
\begin{equation}
	\textsf{E}=\textsf{E}_\text{intra}\cup\textsf{E}_\text{inter}.
\end{equation}
Intracell edges
$\textsf{E}_\text{intra}$
are given by the complete bipartite graph~$K_{LL}$~\cite{dieste2016}
connecting each left vertex to all right vertices and vice versa.
Intercell edges
$\textsf{E}_\text{inter}$
are established by connecting each left vertex of a unit cell to the corresponding left vertex in the cells above and below and by connecting each right vertex to the corresponding right vertex
in the neighboring unit cells left and right.
Edges cross in~$\chi$, so~$\chi$, as well as the variants we consider, are nonplanar graphs.

The connectivity of~$\chi$ affects the efficiency of quantum algorithms implemented on D-Wave computers.
Motivated by technological thrusts,
we study the continuous-time QW on~$\chi$'s variants.
Now we describe~$\chi$ and its variants,
with each considered under two different boundary-conditions indicated by superscripts:
periodic, labelled~${}^\text{p}$, which is equivalent to the graph being on the torus,
and reflecting, labelled~${}^\text{r}$.
Our enhanced chimera graph, denoted by~$\chi^+$,
adds vertical intracell connections:
the top and bottom vertices are connected for~$L=2$ on the left and right side;
the top and bottom vertices are each connected to the middle vertex for~$L=3$ on the left and right side;
and, for~$L\geq4$,
the top left and right are connected to the second-from-top left and right, respectively,
and the bottom left and right are connected to the second-from-bottom left and right, respectively.
We show~$\chi^+$ in Fig.~\ref{fig:Chimeragrid16x16}.

We also introduce the diminished chimera graph~$\chi^-$
resulting from randomly deleting 2\% of the vertices,
which corresponds to the case of real-world quantum annealers with a certain fraction of its vertices not working properly,
and 2\% is the failure rate for the D-Wave 2000Q~\cite{king2017}.
Open-system effects play an important role in how D-Wave quantum computers works
and open-system dynamics can even be beneficial:
quantum annealing could exploit
a thermal environment to speed up compared with closed-system dynamics~\cite{dickson2013}.
Despite potential advantages of open-system dynamics,
our focus is on the closed-system QW on chimera graph as the closed system is advantageous for unitary quantum computing and the chief advantage of the QW analysis is in discovering graph symmetries and effects of localization.

\section{Quantum walk on chimera graph and its variants}
\label{sec:result}
In this section,
we study the continuous-time QW on~$\chi$ and its variants with different boundary conditions and initial positions.
We compare the continuous-time QW and the random walk on~$\chi$.

\begin{figure}
   \includegraphics[width=0.3\columnwidth]{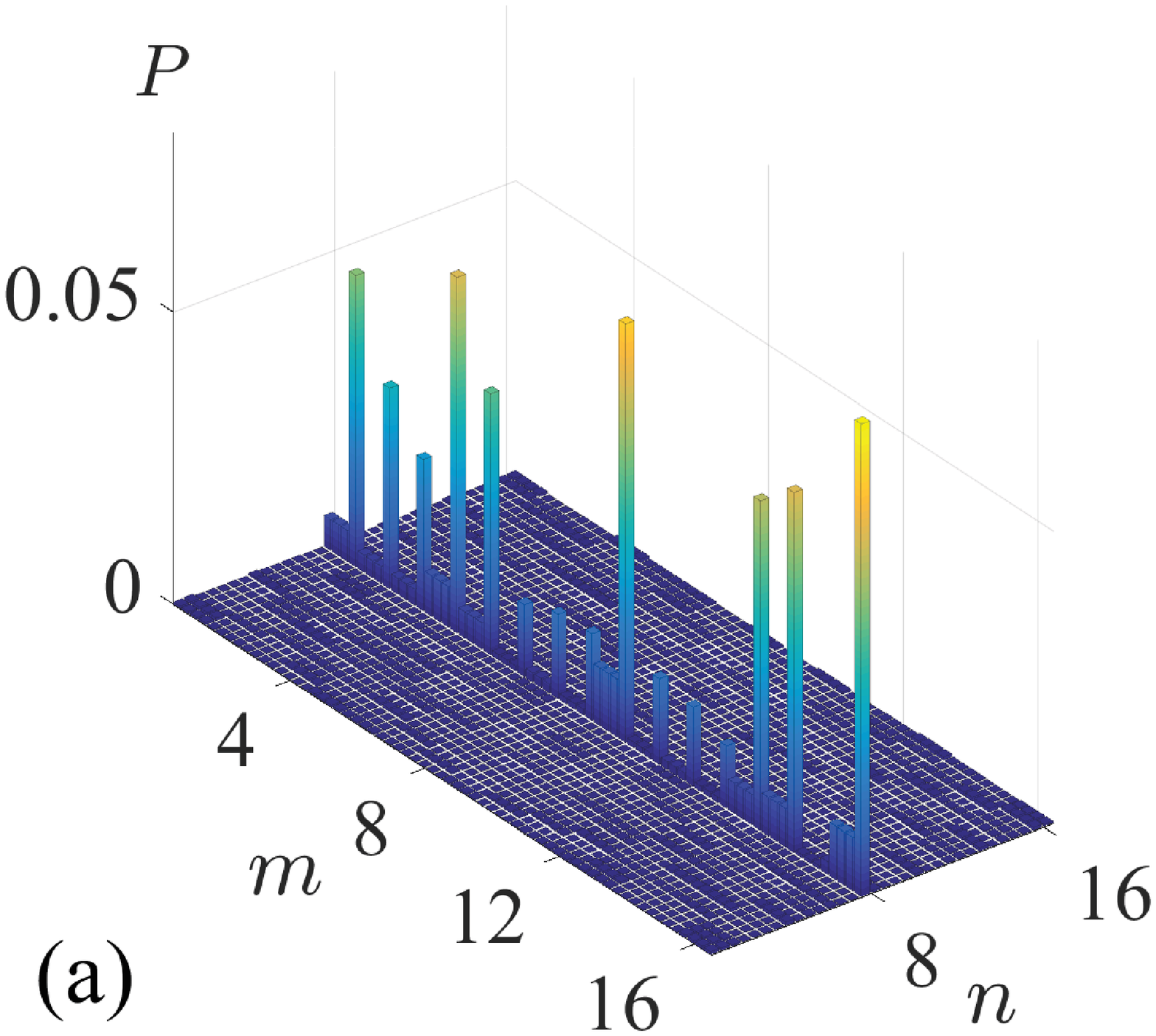} \ \ \
   \includegraphics[width=0.3\columnwidth]{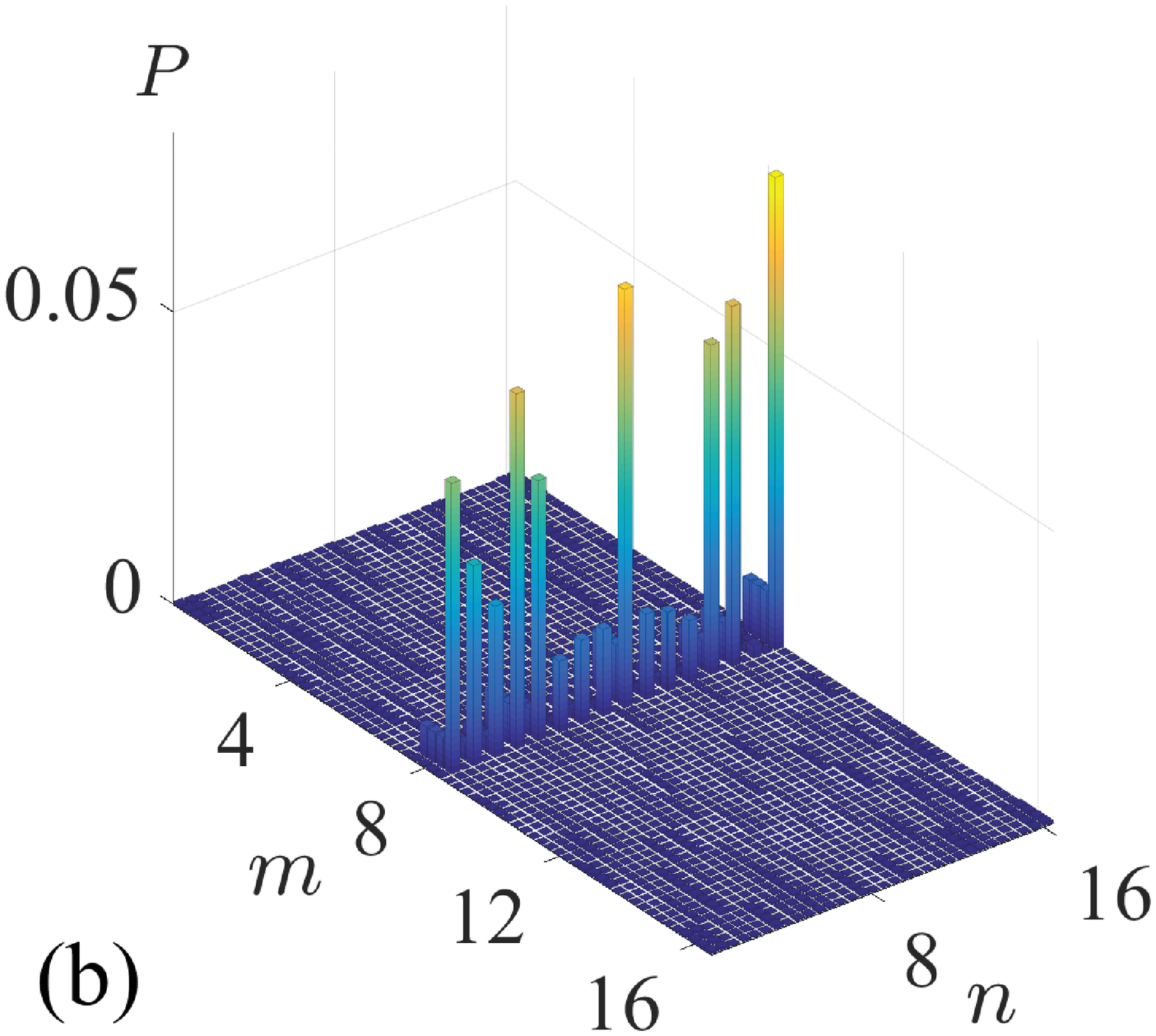} \ \ \
   \includegraphics[width=0.3\columnwidth]{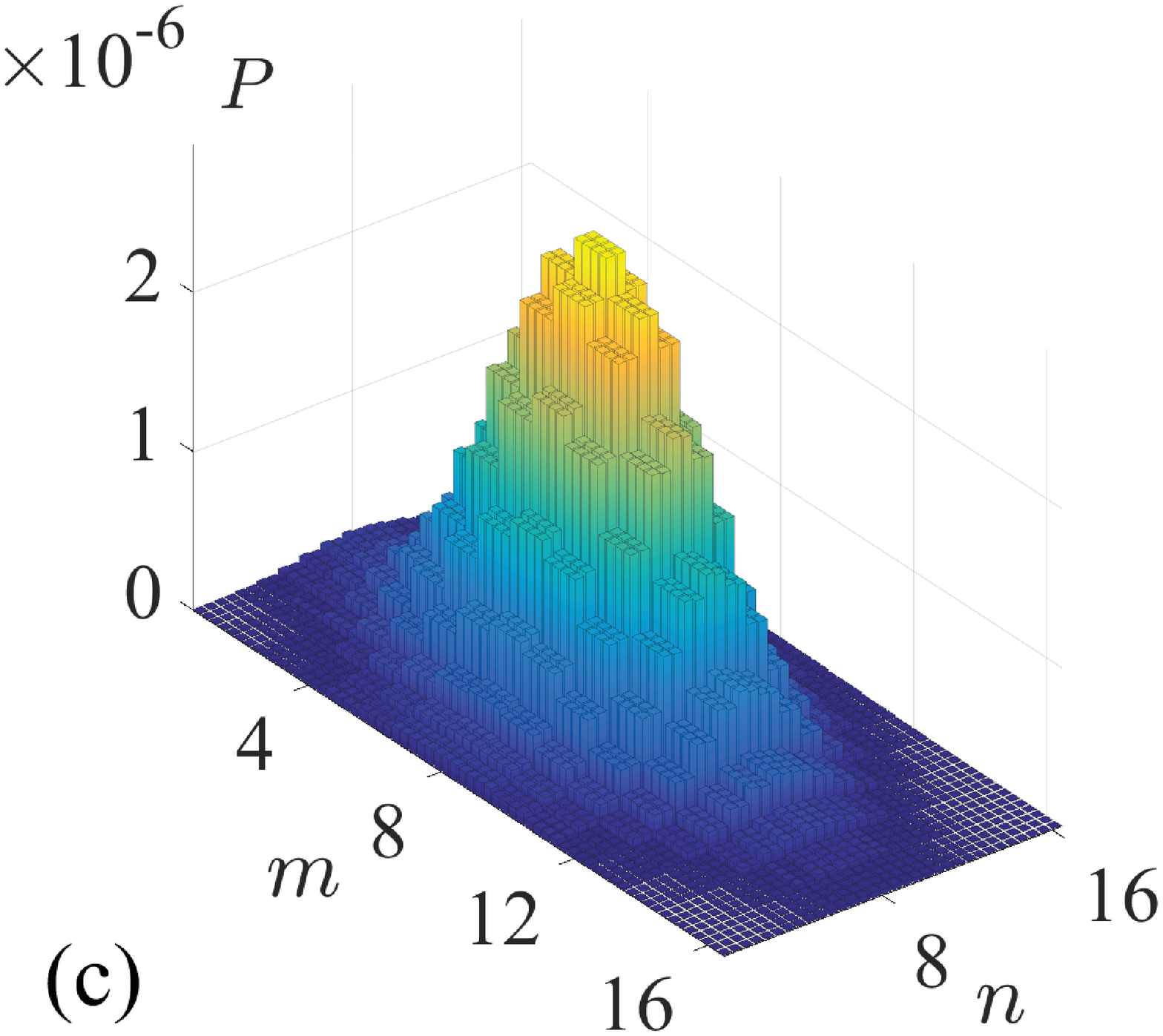} \ \ \
   \includegraphics[width=0.3\columnwidth]{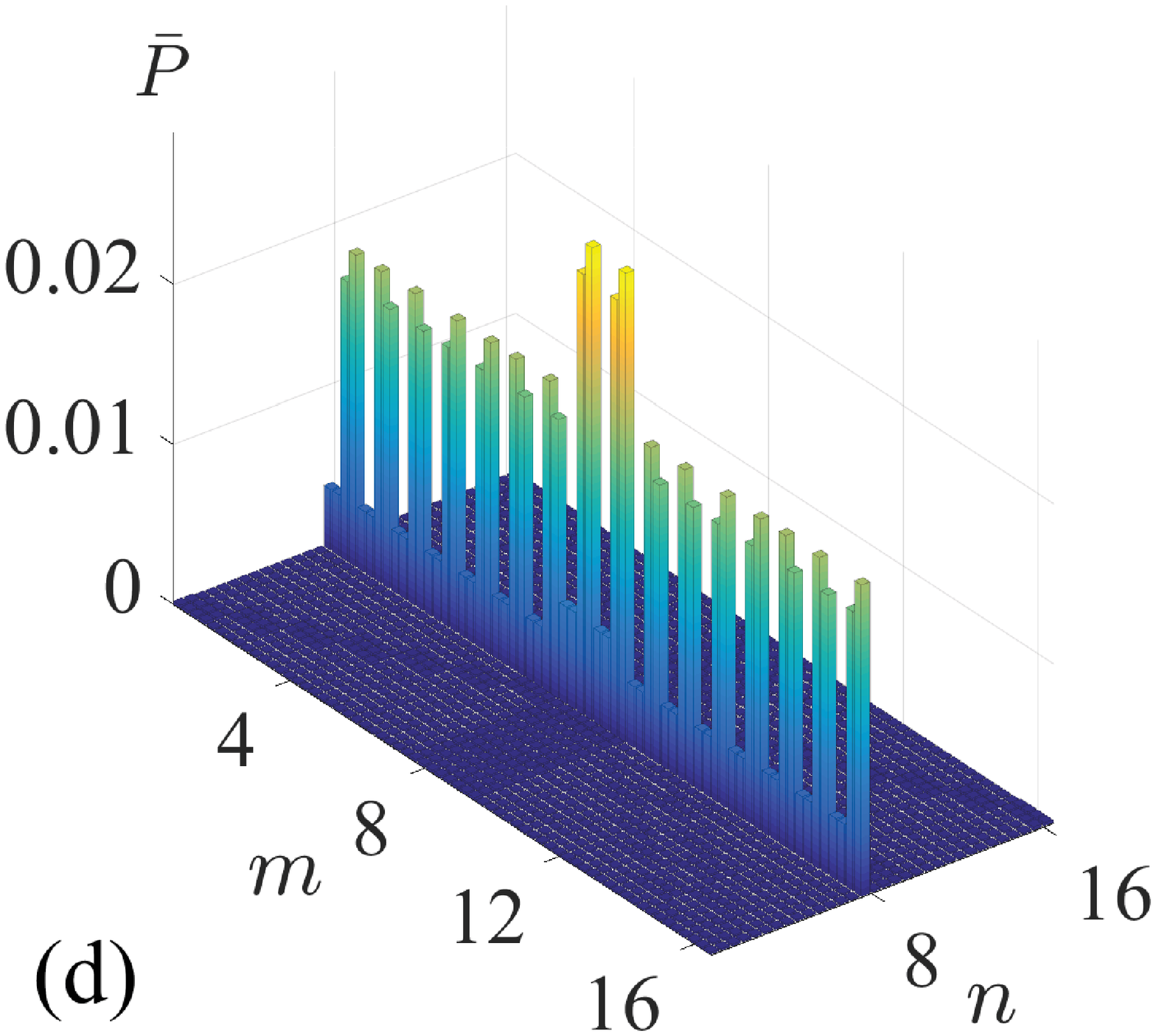} \ \ \
   \includegraphics[width=0.3\columnwidth]{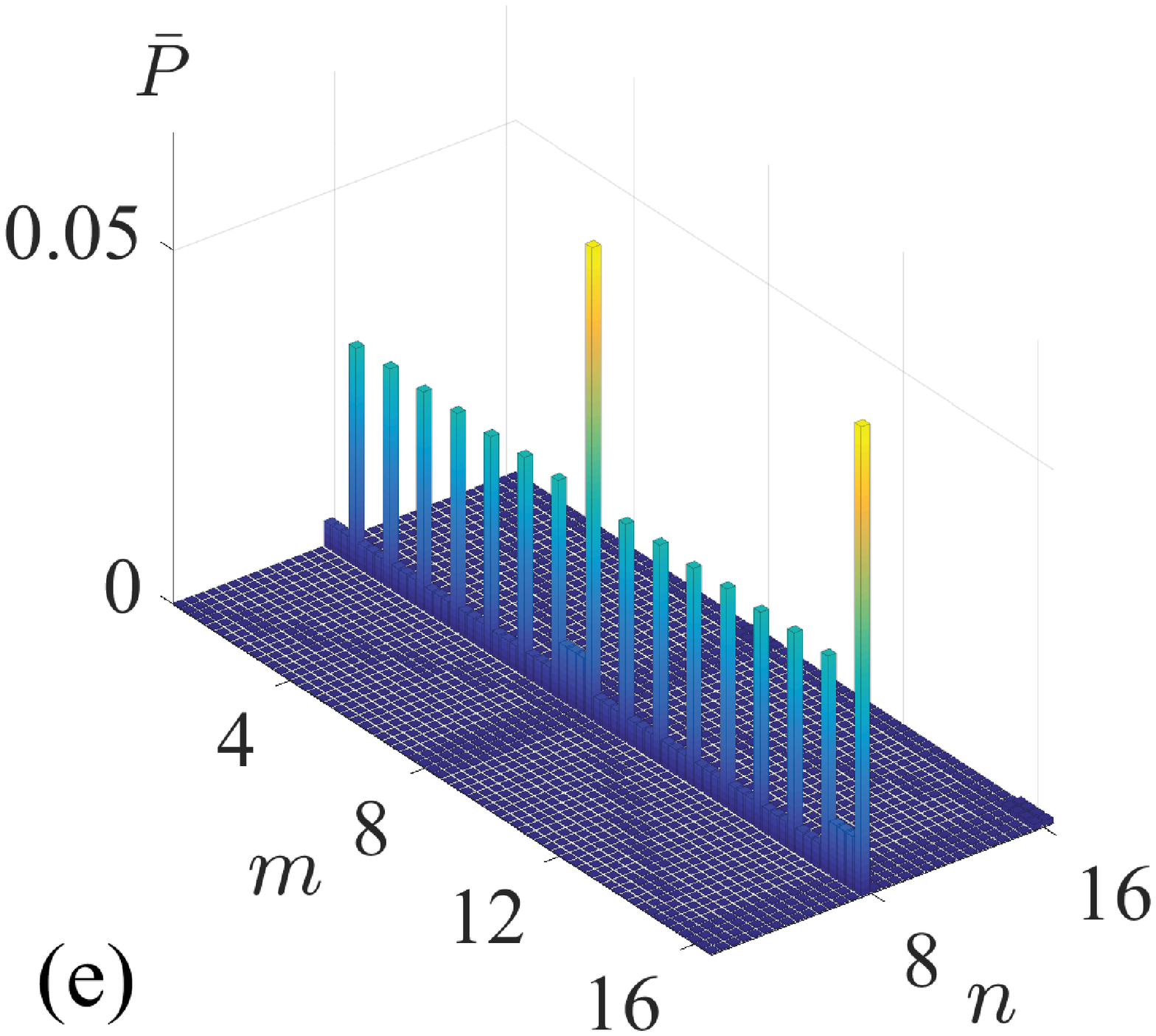} \ \ \
   \includegraphics[width=0.3\columnwidth]{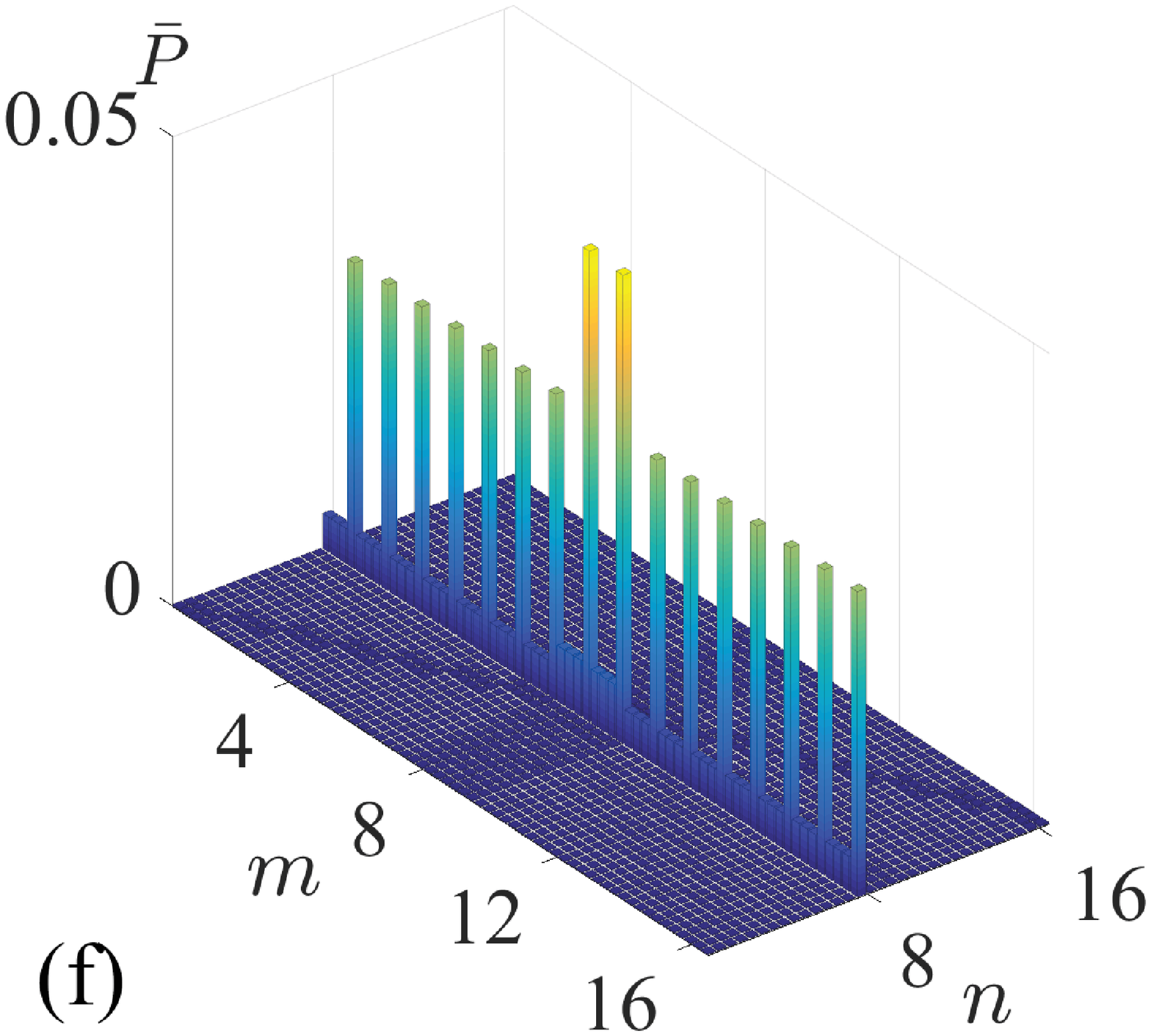}
\caption{%
	 Plots of $P$ for~$\chi_{16,16,4}$ at $t=12$
	and all edges of weight~1
	and intracell label~$\mu$ not shown explicitly.
        (a)~$\chi^{\text{r}}$ initialized at position~$\textsf{v}=(8,8,4)$,
        (b)~$\chi^{\text{r}}$ initialized at position~$\textsf{v}=(8,8,8)$,
        (c)~Classical walk initialized at $\textsf{v}=(8,8,4)$ on $\chi^{\text{r}}$.
	QW limiting distribution for initial position $\textsf{v}=(8,8,4)$
	on (d)~$\chi^{+}$,
        on (e)~$\chi^{\text{p}}$,
	and on (f)~$\chi^{\text{r}}$.%
	}
\label{fig:walkplot}
\end{figure}

First we consider time-independent~$\hat{H}$ with~$j_{\textsf{v}\textsf{v}'}=k_{\textsf{v}\textsf{v}'}=1$ in Eq.~(\ref{eq:walkhamiltonian})
and constrain~$\{h_{\textsf{v}}\}$ by requiring that rows and columns add to zero.
If the walker is initialized at a single vertex~$\textsf{v}$ in the left side of a cell,
we observe that the QW is localized vertically as shown in Fig.~\ref{fig:walkplot}(a),
whereas initialization in the right side of a cell
yields a horizontal localization as shown in Fig.~\ref{fig:walkplot}(b) for the probability distribution
\begin{equation}
	P:=\left\{P_{\textsf{v}\textsf{v}'};\textsf{v}'\in\textsf{V}\right\}
\end{equation}
and the limiting probability distribution
\begin{equation}
	\bar{P}:=\left\{\bar{P}_{\textsf{v}\textsf{v}'};\textsf{v}'\in\textsf{V}\right\}.
\end{equation}
In contrast, the classical random walk is seen to spread in all directions as shown in Fig.~\ref{fig:walkplot}(c).
The classical random walk probability distribution
is for the stochastic walker's likelihood to be at any vertex~\cite{mulken2005}.
The QW exhibits strong localization whereas the classical walk does not;
this localization is evident in other QWs as well~\cite{keating2007,kollar2015,ambainis2016}.
The contrast between periodic and boundary conditions is evident when the walker has evolved to reach a boundary and shows different interference effects,
shown as limiting probability distributions in Figs.~\ref{fig:walkplot}(e,f).

The quantum walker on~$\chi^+$ also shows the similar localization behaviour as on~$\chi^{\text{r}}$,
that is,
if the walker is initialized at a single vertex~$\textsf{v}$ in the left side of a cell,
we observe that the QW is localized vertically as shown in Fig.~\ref{fig:walkplot}(d) for the limiting probability distribution,
whereas initialization in the right side of a cell yields a horizontal localization.
The walker's distribution inside unit cells of~$\chi^+$ is divergent compared with on~$\chi^{\text{r}}$ as shown in Figs.~\ref{fig:walkplot}(d,f),
which reflects the difference between the unit cell structures.

\begin{figure}
  \includegraphics[width=0.3\columnwidth]{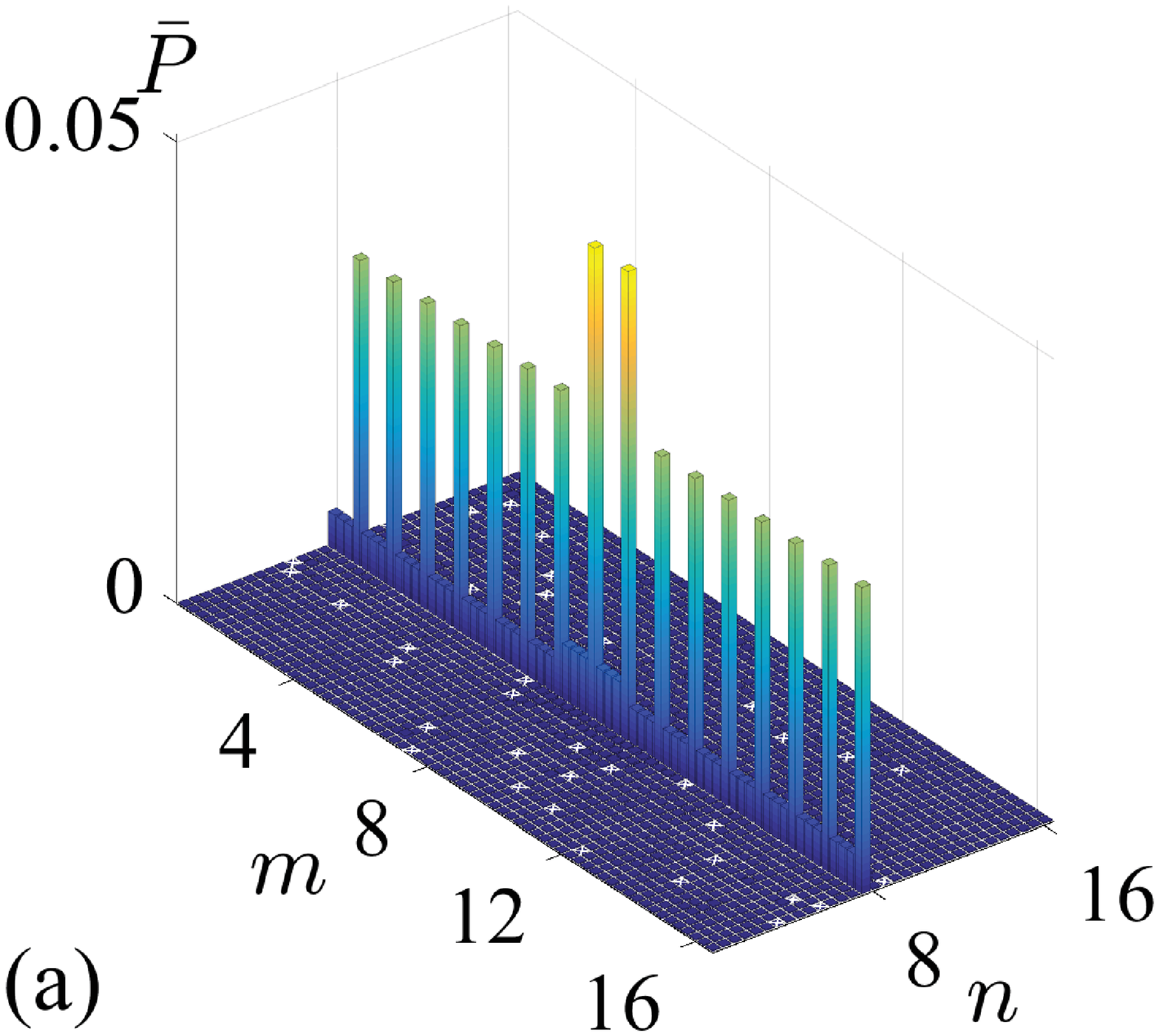} \ \ \ \
   \includegraphics[width=0.3\columnwidth]{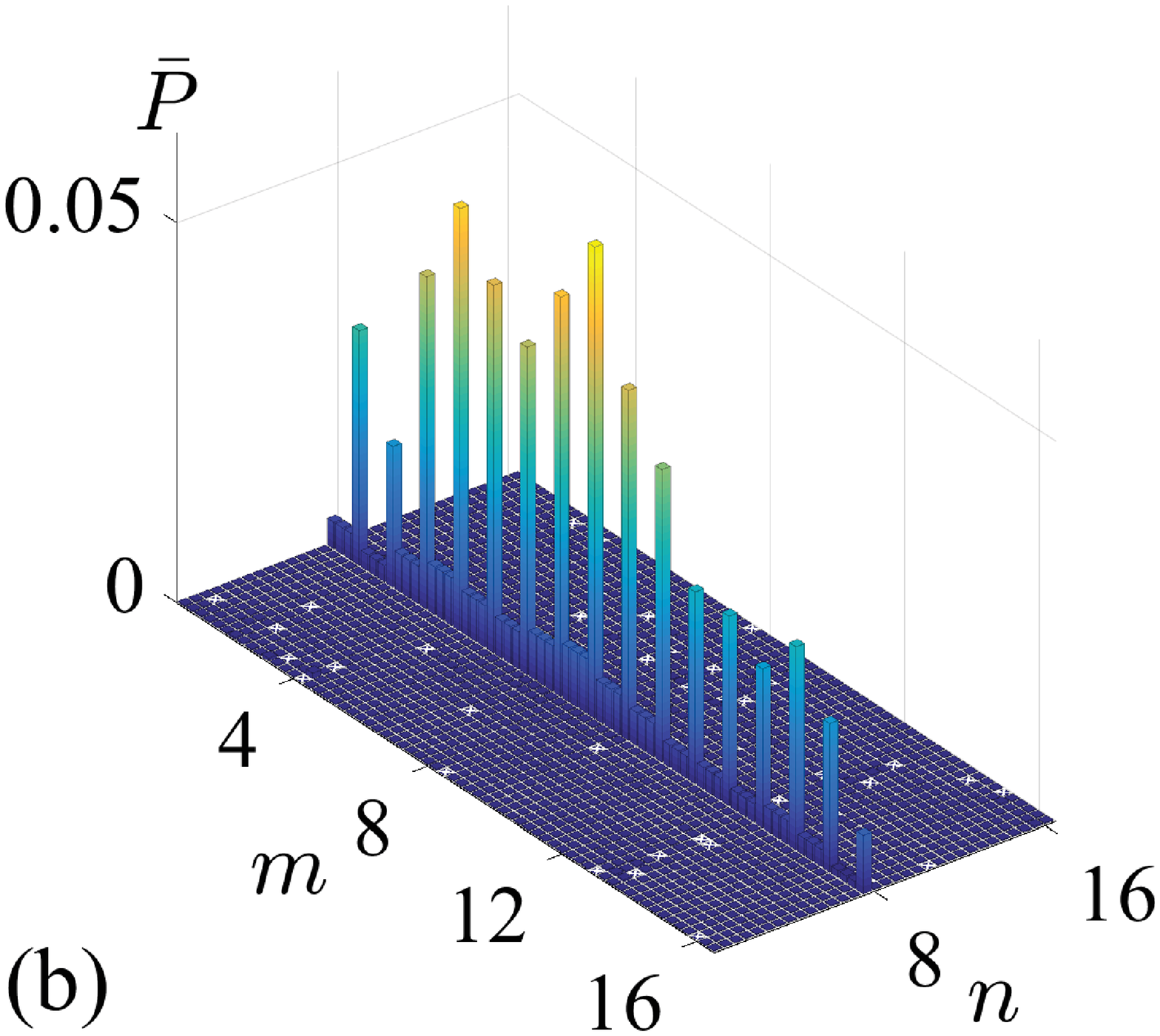}
\caption{%
	QW limiting distribution for initial position~$\textsf{v}=(8,8,4)$. 
	White crosses show the positions of broken vertices.
        Two cases of broken vertices are shown:
        (a) no broken vertex in the 8th row and 8th column and
        (b) at lease one broken vertex located in the 8th row cell or in the 8th column cell.
        }
\label{fig:broken}
\end{figure}

Consideration of~$\chi^-$ is driven by the experimental problem where 2\% vertices do not work~\cite{king2017},
Thus, we consider the class of random~$\chi$ with uniform deletion of 2\% of vertices,
i.e., $\chi^-$ with 2\% of vertices deleted.
We observe that the effect of broken vertices can be divided into two cases.
In the first case, we forbid any broken vertices sharing the same~$m$ and~$n$ as the walker,
whose initial position is~$(m,n,\mu)$.
In this case, broken vertices have little effect on the walker's dynamics,
as seen by comparing Fig.~\ref{fig:broken}(a) to Fig.~\ref{fig:walkplot}(f).
The only noticeable difference is in the low-probability features appearing as dark lines radiating orthogonal to the walker's line of confinement and also diagonally in Fig.~\ref{fig:walkplot}(f). but not in Fig.~\ref{fig:broken}(a).
Quantitatively, we calculate the 1-norm distance between these two limiting probability distributions as
\begin{equation}
	\sum_{\textsf{v},\textsf{v}'} \left|\bar{P}^{\text{r}}_{\textsf{v}\textsf{v}'}
		-\bar{P}^-_{\textsf{v}\textsf{v}'}\right|\approx0.0565.
\end{equation}
In the second case,
we consider at least one broken vertex located with the same~$m$ or~$n$ or both.
In this second case,
the broken vertices have a significant effect on QW behaviour, 
seen by comparing Fig.~\ref{fig:broken}(b) with Fig.~\ref{fig:walkplot}(f).
the distance between those two limiting probability distribution is
\begin{equation}
	\sum_{\textsf{v},\textsf{v}'} \left|\bar{P}^{\text{r}}_{\textsf{v}\textsf{v}'}
		-\bar{P}^-_{\textsf{v}\textsf{v}'}\right|\approx0.3043,
\end{equation}
and the qualitative structure of the curve has been dramatically changed.

\section{Symmetry and spectrum analysis}
\label{sec:symmetries}
From our simulation,
we observe that the quantum walker is strongly localized,
which is related to the structure of the graph.
To analyze and explain the QW localized behaviour,
we now determine the graph symmetries.
Those symmetries manifest as self-adjoint operators~$\hat{\bm{S}}$
that commute with~$\hat{H}$,
i.e.,
\begin{equation}
\label{eq:[SH]=0}
	\left[\hat{\bm{S}},\hat{H}\right]=0\;\forall\hat{\bm{S}}.
\end{equation}
These symmetry operators~$\hat{\bm{S}}$
provide sufficient eigenvalues to
lift the energy degeneracy for eigenstates of~$\hat H$
and thus label all eigenstates uniquely.
In quantum mechanics, 
this set of operators is also known as a complete set of commuting observables.
We construct~$\hat{\bm{S}}$
by determining a sufficient set of symmetries of~$\chi$,
arising from a sufficient generating set
for a subgroup of~$\chi$'s automorphism group,
and we obtain this sufficient generating set.

A graph automorphism~$\alpha(\textsf{G})$
is constructed from a vertex permutation~$\sigma(\textsf{V})$
such that an edge~$(\sigma(\textsf{v}),\sigma(\textsf{v}'))$
exists if and only if~$(\textsf{v},\textsf{v}')$ is an edge in~$\textsf{G}$.
For each automorphism identified by vertex permutation~$\sigma$,
the corresponding symmetry operator is
\begin{equation}
\label{eq:automorphism2operator}
  S_{\textsf{vv}'}
  	=\braket{\textsf{v}|\hat{S}|\textsf{v}'}
	= \braket{\textsf{v}|\sigma\left(\textsf{v}'\right)}
	\in\{0,1\}.
\end{equation}
If matrix~$S$ is not Hermitian,
we replace~$S$ by
\begin{equation}
\label{eq:herm}
	\operatorname{herm}S:=\frac{1}{2}\left(S+S^\dagger\right),
\end{equation}
for~$S^\dagger$ the adjoint
so we discuss~$\hat{S}$ operators as being self-adjoint below.

\subsection{Symmetry and spectrum analysis for~$\chi_{MNL}^{\text{p}}$}
\label{subsec:chip}
In this subsection we study the symmetries of the periodic chimera graph~$\chi^{\text{p}}$.
Figure~\ref{fig:Chimeragrid16x16} shows the symmetries of $\chi_{16,16,4}^{\text{p}}$.
Specifically,
$\sigma_{1,3}$ is an intracell left-side (right-side) translation permutation symmetry,
$\sigma_{2,4}$ is an intracell left-side (right-side) mirror permutation symmetry,
$\sigma_{5,7}$ is a translation permutation along the~$M$~($N$) axis, and
$\sigma_{6,8}$ is a mirror permutation through the~$M$~($N$) axis.
The symmetry operators are
\begin{equation}
\label{eq:S}
	\hat{\bm{S}}
		=\left(\hat{S}_1, \hat{S}_2, \dots, \hat{S}_8\right),
\end{equation}
with matrices~$S_i$ obtained from Eq.~(\ref{eq:automorphism2operator})
and satisfying Eq.~(\ref{eq:[SH]=0}).
Each~$S_i$
is (not) Hermitian for even (odd)~$i$
and replaced by~$\operatorname{herm}S_i$ for odd~$i$.
Table~$\ref{tab:sigma}$ shows the detail of~$\sigma_{i}$ for~$\chi_{MNL}^{\text{p}}$.
The first column shows the vertex permutation;
the second column shows the vertex coordinate defined in Eq.~(\ref{eq:VMNL});
the third column shows the vertex coordinate after permutation;
the fourth column gives the eigenvalues of~$S$ corresponding to~$\sigma$ given by Eq.~(\ref{eq:automorphism2operator}).
\begin{table} \centering
\caption{The permutation operators resulted from graph automorphism for~$\chi_{MNL}^{\text{p}}$}
\label{tab:sigma}
\begin{tabular}{| c | c | l | c |}\hline
   Vertex permutation   & $\textsf{v}$   &\multicolumn{1}{c |}{$\sigma(\textsf{v})$} & Eigenvalues of $S$ \\ \hline
$\sigma_{1}$ & $(m,n,\mu)$ & \tabincell{l}{$(m,n,\mu+1), \mu \in [1,L-1]$ \\ $(m,n,1), \mu=L$ \\ $(m,n,\mu), \mu \in [L+1,2L]$} & $\cos(\frac{2\pi\imath}{L}), \imath \in [0,L-1]$ \\ \hline
$\sigma_{2}$ & $(m,n,\mu)$ & \tabincell{l}{$(m,n,L+1-\mu), \mu \in [1,L]$ \\ $(m,n,\mu),\mu \in [L+1,2L]$} & $-1,1$ \\ \hline
$\sigma_{3}$ & $(m,n,\mu)$ & \tabincell{l}{$(m,n,\mu), \mu \in [1,L]$ \\ $(m,n,\mu+1), \mu \in [L+1,2L-1]$ \\ $(m,n,L+1), \mu=2L$} & $\cos(\frac{2\pi\imath'}{L}), \imath' \in [0,L-1]$ \\ \hline
$\sigma_{4}$ & $(m,n,\mu)$ & \tabincell{l}{$(m,n,\mu), \mu \in [1,L]$ \\ $(m,n,2L+1-\mu), \mu=[L+1,2L]$} & $-1,1$ \\ \hline
$\sigma_{5}$ & $(m,n,\mu)$ & \tabincell{l}{$(m+1,n,\mu), m \in [1,M-1]$ \\ $(1,n,\mu), m=M$} & $\cos(\frac{2\pi\jmath}{L}), \jmath \in [0,M-1]$ \\ \hline
$\sigma_{6}$ & $(m,n,\mu)$ & $(M+1-m,n,\mu)$ & $-1,1$ \\ \hline
$\sigma_{7}$ & $(m,n,\mu)$ & \tabincell{l}{$(m,n+1,\mu), n \in [1,N-1]$ \\ $(m,1,\mu), n=N$} & $\cos(\frac{2\pi\jmath'}{L}), \jmath' \in [0,N-1]$ \\ \hline
$\sigma_{8}$ & $(m,n,\mu)$ & $(m,N+1-n,\mu)$ & $-1,1$ \\ \hline
\end{tabular}
\end{table}

To analyze and explain the QW localized behaviour,
we construct the spectrum and stationary states.
Stationary states are~$\hat{H}$ eigenvectors,
with each uniquely labeled by energy~$E$
and spectrum
\begin{equation}
	\bm{s}\in\operatorname{spec}\hat{\bm S}\subset\mathbb{R}^8:
		\hat{H}\ket{E\bm{s}}=E\ket{E\bm{s}}.
\end{equation}
We parameterize $\operatorname{spec}\hat{\bm S}$
by unit-lattice coordinates
\begin{equation}
\label{eq:spectrallabels}
\bm{\imath}:=\left(\imath, \imath', \jmath,\jmath'\right)
		 \in\mathbb{Z}^4
\end{equation}
expressed as a disjoint union of two three-dimensional unit lattices and one two-dimensional lattices.
For
\begin{equation}
	[N]:=\{0,\dots,N\},
\end{equation}
these two three-dimensional lattices are
\begin{align}
\mathbb X_{\triangleright}&:=[L-1]\backslash\{0\}\times\{0\}\times[M-1]\times[N-1],\label{eq:xleft} \\
\mathbb X_{\triangleleft}&:=\{0\}\times[L-1]\backslash\{0\}\times[M-1]\times[N-1]
\end{align}
and
\begin{equation}
	\mathbb X_{\bowtie}:=\{0\}\times\{0\}\times[M-1]\times[N-1]
\end{equation}
is the two-dimensional unit lattice.

Spectral labels are expressed as $\bm{s}(\bm{\imath})$
and~$E(\bm{\imath})$.
For a valid~$\bm{\imath}$,
\begin{align}
	s_1=&\cos\frac{2\pi\imath}{L},\;
	s_2=\begin{cases}
		1,&\imath<\frac{L}{2},\\
		-1,&\imath\geq\frac{L}{2},
		\end{cases} \label{eq:s1s2} \\
	s_3=&\cos\frac{2\pi\imath'}{L},\;
	s_4=\begin{cases}
		1,&\imath'<\frac{L}{2},\\
		-1,&\imath'\geq\frac{L}{2},
		\end{cases} \label{eq:s3s4} \\
	s_5=&\cos\frac{2\pi\jmath}{M},\;
	s_6=\begin{cases}
		1,&\jmath<\frac{M}{2},\\
		-1,&\jmath\geq\frac{M}{2},
		\end{cases}\\
	s_7=&\cos\frac{2\pi\jmath'}{N},\;
	s_8=\begin{cases}
		1,&\jmath'<\frac{N}{2},\\
		-1,&\jmath'\geq\frac{N}{2},
		\end{cases}
\end{align}
and the energy is
\begin{equation}
\label{eq:eigenvalueE}
	E=\begin{cases}
		L+2-2s_5,&\imath\neq0,\imath'=0,\\
		L+2-2s_7,&\imath=0,\imath'\neq0,\\
		(L+2)-(s_5+s_7)\pm\sqrt{L^2+(s_5-s_7)^{2}},
			&\imath=0,\imath'=0.
	    \end{cases}
\end{equation}
The cardinalities are
\begin{equation}
\label{eq:cardinalities}
	\left|\mathbb X_{\triangleright}\right|=MN(L-1)=\left|\mathbb X_{\triangleleft}\right|,\;
	\left|\mathbb X_{\bowtie}\right|=2MN.
\end{equation}
In this subsection, we have obtained the spectrum for the~$\chi^{\text{p}}$ QW Hamiltonian.
Also we divided the set of stationary states into three cases for interpretation in~\S\ref{sec:discussion}.

\subsection{Symmetry and spectrum analysis for~$\chi_{MNL}^{\text{r}}$}
\label{subsec:chir}
In this subsection we study the symmetries of the reflecting chimera graph~$\chi^{\text{r}}$.
Both $\chi_{MNL}^{\text{r}}$ and~$\chi_{MNL}^{\text{p}}$ have identical intracell symmetries; i.e.,
for~$\chi_{MNL}^{\text{r}}$,
the vertex permutation symmetries $\sigma_{1,2,3,4}$ still hold.
However, due to reflecting boundary conditions,
translation permutation symmetry along the~$M$~($N$) axis no longer holds in the reflecting case.
In order to lift the intercell degeneracy of the QW~$\hat{H}$,
we need to construct other operators.
Notice that the QW~$\hat{H}$ of the one-dimensional finite line has no degeneracy,
so we write the matrix representation of~$S'_{5,6}$ directly as
\begin{equation}
S'_{5}=A_M\otimes\mathbb{1}_{N}\otimes\mathbb{1}_{2L},\  S'_{6}=\mathbb{1}_M\otimes A_{N}\otimes\mathbb{1}_{2L},
\end{equation}
with~$A_M$ the QW~$\hat{H}$ for the one-dimensional finite line with~$M$ vertices.

The elements of~$A_M$ are
\begin{equation}
	A_{ij}
		=\begin{cases}
			-1,&|i-j|=1,\\
			2,&i=j,i\neq1,i\neq M, \\
			1,&i=j,i=1,i=M, \\
			0,&\text{otherwise,}\\
        \end{cases}
\end{equation}
and the eigenvalues of~$S'_{5,6}$ are
\begin{align}
	s'_5=&2+2\cos\frac{\pi\jmath}{M},\;\jmath\in[M]\backslash\{0\},
		\nonumber\\
	s'_6=&2+2\cos\frac{\pi\jmath'}{N},\;\jmath'\in[N]\backslash\{0\}.
\end{align}
Thus, for~$\chi_{MNL}^{\text{r}}$, the symmetry operators are
\begin{equation}
\hat{\bm{S}}
	=\left(\hat{S}_1, \hat{S}_2, \dots, \hat{S}'_5,\hat{S}'_6\right),
\end{equation}
with matrices~$S_i$ obtained from Eq.~(\ref{eq:automorphism2operator}).

Stationary states are~$\hat{H}$ eigenvectors,
with each uniquely labeled by energy~$E$
and spectrum
\begin{equation}
\label{eq:uniquelylabeled+}
	\bm{s}\in\operatorname{spec}\hat{\bm S}\subset\mathbb{R}^6:
		\hat{H}\ket{E\bm{s}}=E\ket{E\bm{s}}.
\end{equation}
We parameterize $\operatorname{spec}\hat{\bm S}$
by unit-lattice coordinates
\begin{equation}
\bm{\imath}:=\left(\imath, \imath', \jmath,\jmath'\right)
		 \in\mathbb{Z}^4
\end{equation}
expressed as a disjoint union of two three-dimensional unit lattices
\begin{align}
\mathbb X_{\triangleright}&:=[L-1]\backslash\{0\}\times\{0\}\times[M]\backslash\{0\}\times[N]\backslash\{0\},\\
\mathbb X_{\triangleleft}&:=\{0\}\times[L-1]\backslash\{0\}\times[M]\backslash\{0\}\times[N]\backslash\{0\}
\end{align}
and one two-dimensional unit lattice
\begin{equation}
\mathbb X_{\bowtie}:=\{0\}\times\{0\}\times[M]\backslash\{0\}\times[N]\backslash\{0\}.
\end{equation}

Spectral labels are expressed as $\bm{s}(\bm{\imath})$
and~$E(\bm{\imath})$.
For a valid~$\bm{\imath}$,
\begin{align}
	s_1=&\cos\frac{2\pi\imath}{L},\;
	s_2=\begin{cases}
		1,&\imath<\frac{L}{2},\\
		-1,&\imath\geq\frac{L}{2},
		\end{cases} \label{eq:s1s2} \\
	s_3=&\cos\frac{2\pi\imath'}{L},\;
	s_4=\begin{cases}
		1,&\imath'<\frac{L}{2},\\
		-1,&\imath'\geq\frac{L}{2},
		\end{cases} \label{eq:s3s4} \\
	s'_5=&2+2\cos\frac{\pi\jmath}{M},\;
	s'_6=2+2\cos\frac{\pi\jmath'}{N},
\end{align}
and the energy is
\begin{equation}
\label{eq:eigenvalueEr}
	E=\begin{cases}
		L+s'_5,&\imath\neq0,\imath'=0,\\
		L+s'_6,&\imath=0,\imath'\neq0,\\
		L+(s'_5+s'_6)/2\pm\sqrt{L^2+(s'_5-s'_6)^{2}/4},
			&\imath=0,\imath'=0.
	    \end{cases}
\end{equation}
The cardinalities are the same as Eq.~(\ref{eq:cardinalities}).
In this subsection, we have obtained the spectrum of the~$\chi^{\text{r}}$ QW Hamiltonian.
Furthermore we divided the set of stationary states into three cases for interpretation in \S\ref{sec:discussion}.

\subsection{Symmetry and spectrum analysis for~$\chi_{MNL}^{+}$}
\label{subsec:chi+}
In this subsection we study the symmetries of the enhanced chimera graph~$\chi^{+}$.
For~$\chi_{MNL}^{+}$,
intracell translation permutation symmetry does not exist.
The following demonstration is with~$L=4$ as example.

\begin{figure}
   \includegraphics[width=0.8\columnwidth]{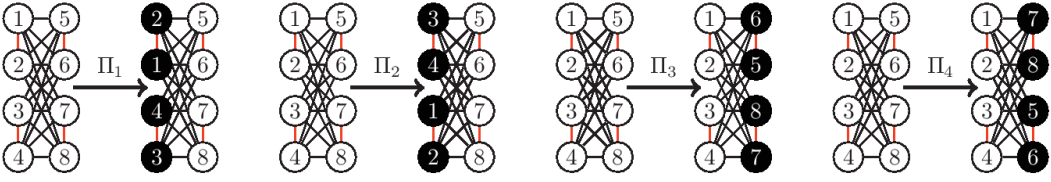} 
\caption{%
	Intracell symmetries of~$\chi_{MNL}^+$ for~$L=4$.%
	}
\label{fig:pi}
\end{figure}

Figure~\ref{fig:pi} shows the intracell symmetries of $\chi_{MNL}^{+}$.
$\Pi_{1,3}$ is an intracell left-side (right-side) permutation symmetry that permutes intracell vertices $1$ and $2$ ($5$ and $6$)
and simultaneously permutes intracell vertices $3$ and $4$ ($7$ and $8$).
$\Pi_{2,4}$ is an intracell left-side (right-side) permutation symmetry that permutes intracell vertices $1$ and $3$ ($5$ and $7$) 
and simultaneously permutes intracell vertices $2$ and $4$ ($6$ and $8$).
Intercell symmetries are the same as for~$\chi_{MNL}^{\text{r}}$,
which are~$S'_{5,6}$.
The symmetry operators are
\begin{equation}
\label{eq:S}
\hat{\bm{S}}
	=\left(\hat{\Pi}_1,\dots,\hat{\Pi}_4,\hat{S}'_5,\hat{S}'_6\right),
\end{equation}
with matrices~$S_i$ obtained from Eq.~(\ref{eq:automorphism2operator}).

Stationary states are~$\hat{H}$ eigenvectors,
with each uniquely labeled by energy~$E$
and spectrum
\begin{equation}
	\bm{s}\in\operatorname{spec}\hat{\bm S}\subset\mathbb{R}^6:
		\hat{H}\ket{E\bm{s}}=E\ket{E\bm{s}}.
\end{equation}
We parameterize $\operatorname{spec}\hat{\bm S}$
by unit-lattice coordinates
\begin{equation}
\bm{\imath}:=\left(\pi_1, \pi_2,\pi_3,\pi_4, \jmath,\jmath'\right)
		 \in\mathbb{Z}^6
\end{equation}
expressed as a disjoint union of two four-dimensional unit lattices
\begin{align}
	\mathbb X_{\triangleright}&:=\{\pm1\}\times\{\pm1\}\times\{1\}\times\{1\}\times[M]\backslash\{0\}\times[N]\backslash\{0\}, \pi_1+\pi_2\neq2,\\
	\mathbb X_{\triangleleft}&:=\{1\}\times\{1\}\times\{\pm1\}\times\{\pm1\}\times[M]\backslash\{0\}\times[N]\backslash\{0\},\pi_3+\pi_4\neq2,
\end{align}
and one two-dimensional unit lattice
\begin{equation}
	\mathbb X_{\bowtie}:=\{1\}\times\{1\}\times\{1\}\times\{1\}\times[M]\backslash\{0\}\times[N]\backslash\{0\}.
\end{equation}

Spectral labels are expressed as $\bm{s}(\bm{\imath})$
and~$E(\bm{\imath})$.
For a valid~$\bm{\imath}$,
\begin{align}
	\pi_1=&\pm1,\;
	\pi_2=\pm 1,\;
	\pi_3=\pm 1,\;
	\pi_4=\pm 1,\; \\
	s'_5=&2+2\cos\frac{\pi\jmath}{M},\;
	s'_6=2+2\cos\frac{\pi\jmath'}{N},
\end{align}
and the energy is
\begin{equation}
\label{eq:eigenvalueE+}
	E=\begin{cases}
		L+s'_5,&\pi_1=1,\pi_2=-1,\pi_3=\pi_4=1,\\
		L+2+s'_5,&\pi_1=-1,\pi_2=\pm1,\pi_3=\pi_4=1,\\
		L+s'_6,&\pi_1=\pi_2=1,\pi_3=1,\pi_4=-1,\\
		L+2+s'_6,&\pi_1=\pi_2=1,\pi_3=-1,\pi_4=\pm1,\\
		L+(s'_5+s'_6)/2\pm\sqrt{L^2+(s'_5-s'_6)^{2}/4},
			&\pi_1=\pi_2=\pi_3=\pi_4=1.
	    \end{cases}
\end{equation}
The cardinalities are the same as Eq.~(\ref{eq:cardinalities}).
In this subsection, we obtained the spectrum of the~$\chi^+$ QW Hamiltonian.
Furthermore we divided the set of stationary states into three cases for interpretation in~\S\ref{sec:discussion}.
For other~$L$ values, one could consider symmetry operators such as
\begin{equation}
	\hat{\bm{S}}
		=\left(\hat{\Pi}_1,\dots,\hat{\Pi}_4,\hat{S}_1,\dots,\hat{S}_4,\hat{S}'_5,\hat{S}'_6\right),
\end{equation}
but we do not study these because the method to lift the degeneracy is similar.

\subsection{Spectrum analysis for~$\chi^-$}
\label{subsec:chi-}
In this subsection we study the spectrum for the diminished chimera graph~$\chi^-$.
Symmetry analysis for~$\chi^-$ is difficult as most symmetry operators fail to commute with the Hamiltonian corresponding to a graph with broken vertices.
But we can compare~$\chi^-$ with~$\chi^{\text{r}}$ which we already study above.

\begin{figure}    
\includegraphics[width=0.9\columnwidth]{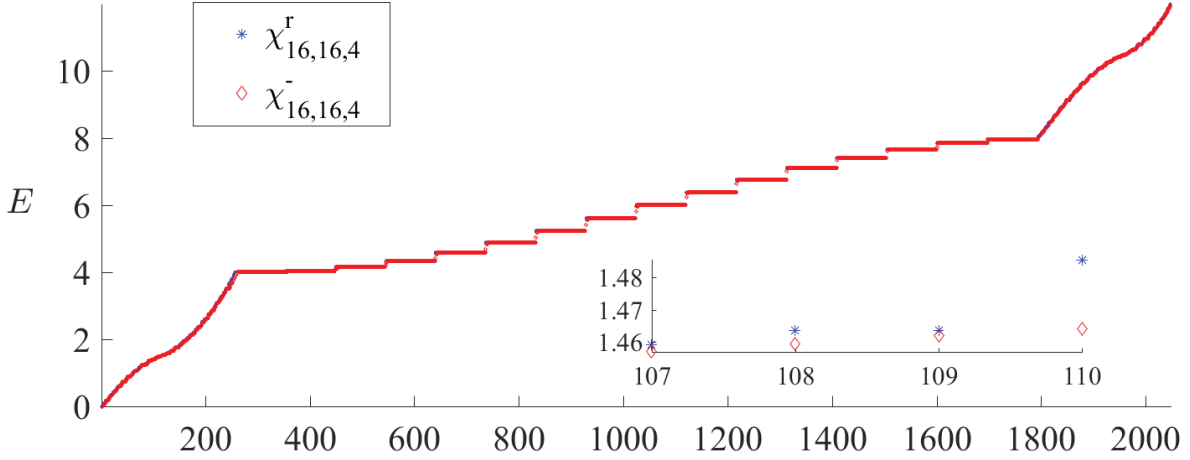} 
\caption{%
	Spectra for~$\chi_{16,16,4}^{\text{r}}$ ({\blue{*}})
	and for~$\chi_{16,16,4}^-$ ({\red{$\diamond$}})
	as eigenvalue vs index for one broken vertex at~$(1,1,1)$
	with the full spectrum shown and the inset showing only the interesting comparative region near $E\approx1.4635$.
	In the interactive online version,
	the figure zooms in from showing the entire spectrum to the restricted spectrum from $E\approx1.4570$
	to~$E\approx1.4850$,
	which shows clearly the degeneracy of $E\approx1.4635$ being 2 for~$\chi_{16,16,4}^{\text{r}}$
	and being nondegenerate for~$\chi_{16,16,4}^-$
	at $E\approx1.4620$ and the other at~$E\approx1.4640$.%
	}
\label{fig:brokeneigen}
\end{figure}

First, we consider the case of just one broken vertex as the simplest example of broken vertices in the graph,
and we compare the spectrum for~$\hat H$ for the unbroken reflecting graph~$\chi_{16,16,4}^{\text{r}}$
vs the reflecting graph with one broken vertex, namely~$\chi_{16,16,4}^-$.
The two spectra are compared in Fig.~\ref{fig:brokeneigen}
with the entire spectrum shown as a zoom-in plot online
and as a full-spectrum plot with partial spectrum plot as an inset.
By zooming in,
we see that the unbroken reflecting graph has a doubly degenerate eigenvalue~$E=1.4635$,
which reduces to nondegeneracy arising from the broken vertex.
This nondegeneracy is manifested
as one eigenvalue shifting to a slightly lower value~$E\approx1.4620$
and the other eigenvalue moving to the slightly higher $E\approx1.4640$.
This broken vertex only noticeably changes the degeneracy of this eigenvalue and leaves most other eigenvalue degeneracies intact,
thereby showing how only a few stationary states are affected by breaking one vertex.

\begin{figure}   
\includegraphics[width=0.35\columnwidth]{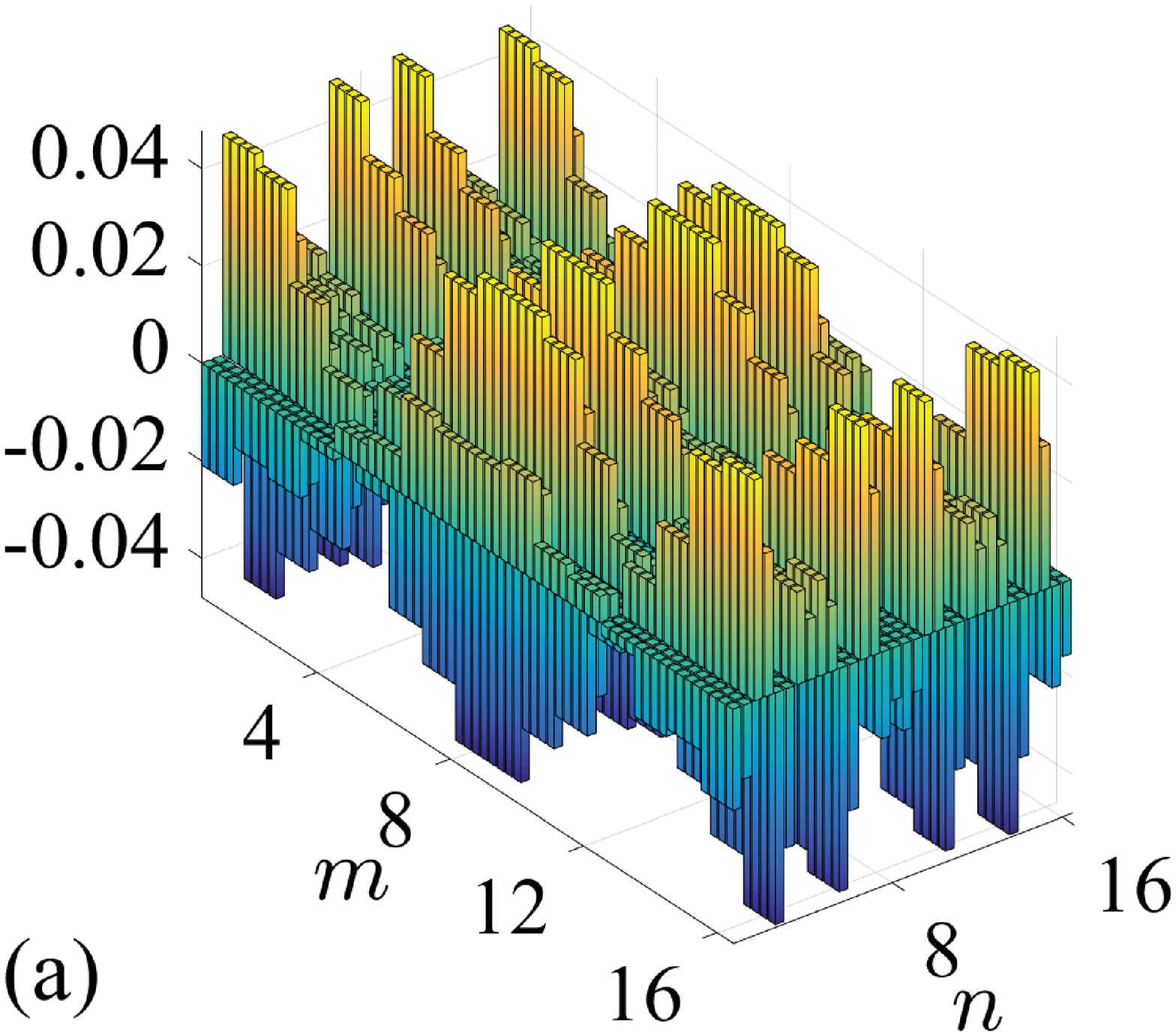} \ \
\includegraphics[width=0.35\columnwidth]{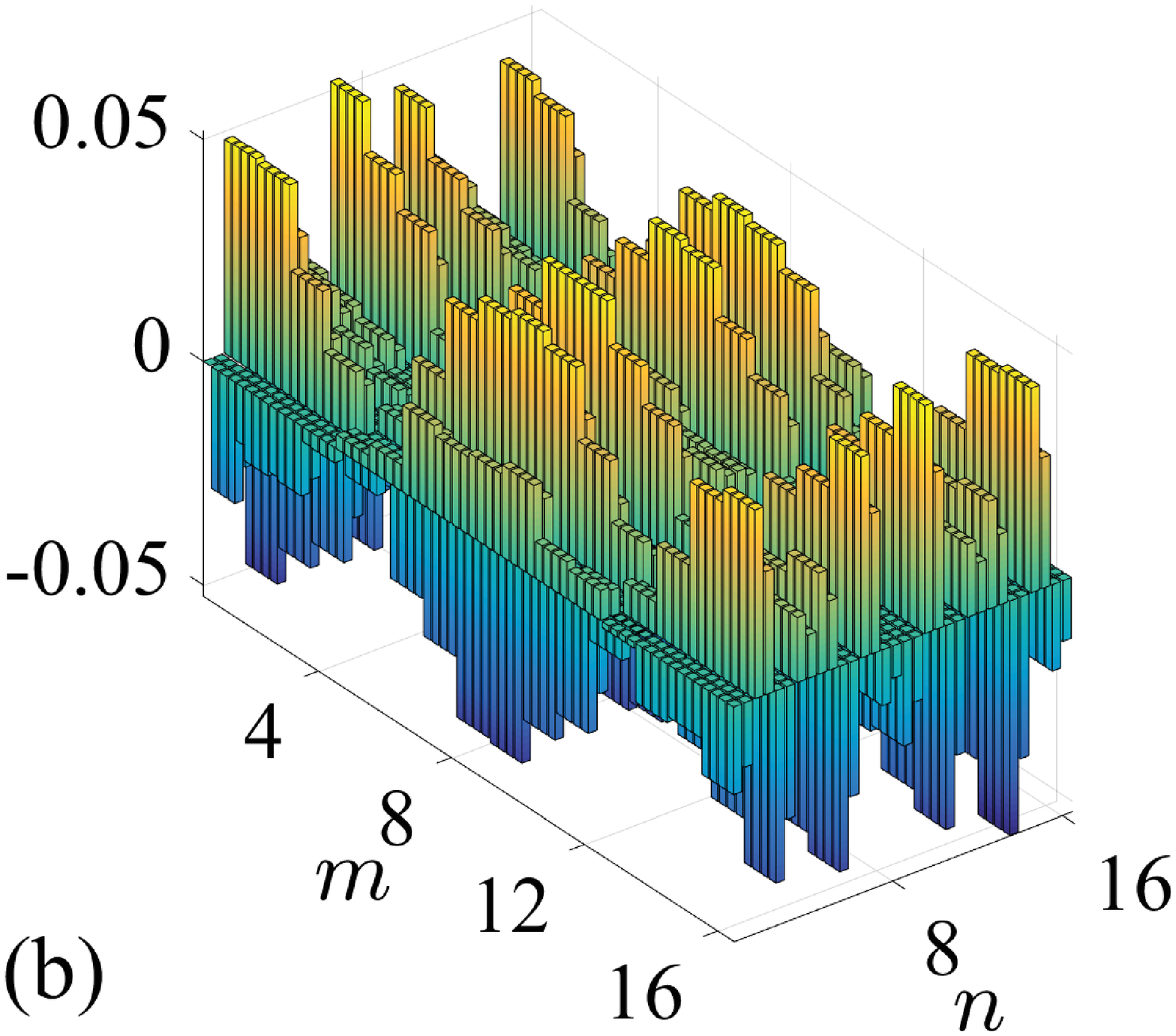}  ~\\
\includegraphics[width=0.35\columnwidth]{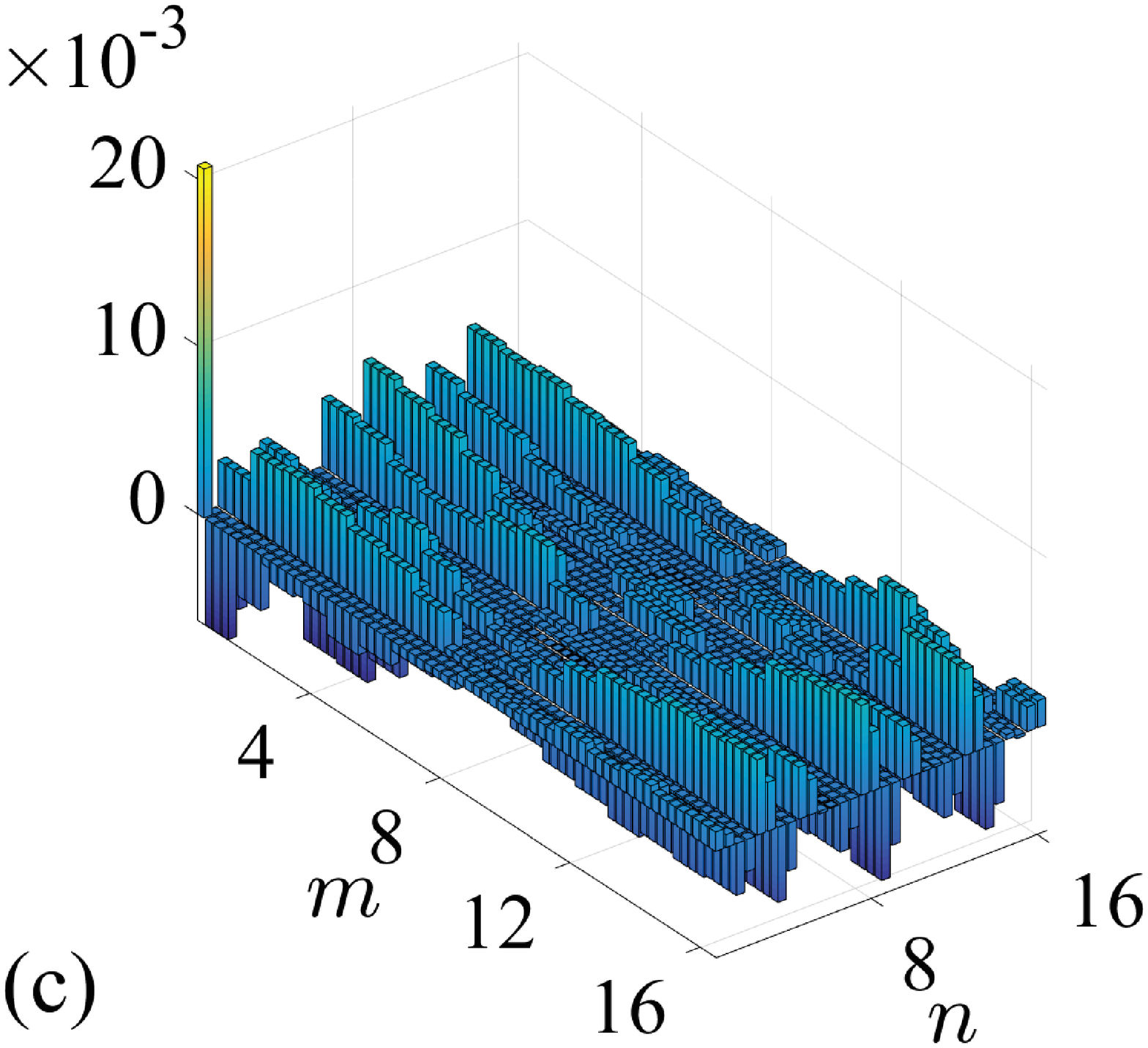} \ \
\includegraphics[width=0.35\columnwidth]{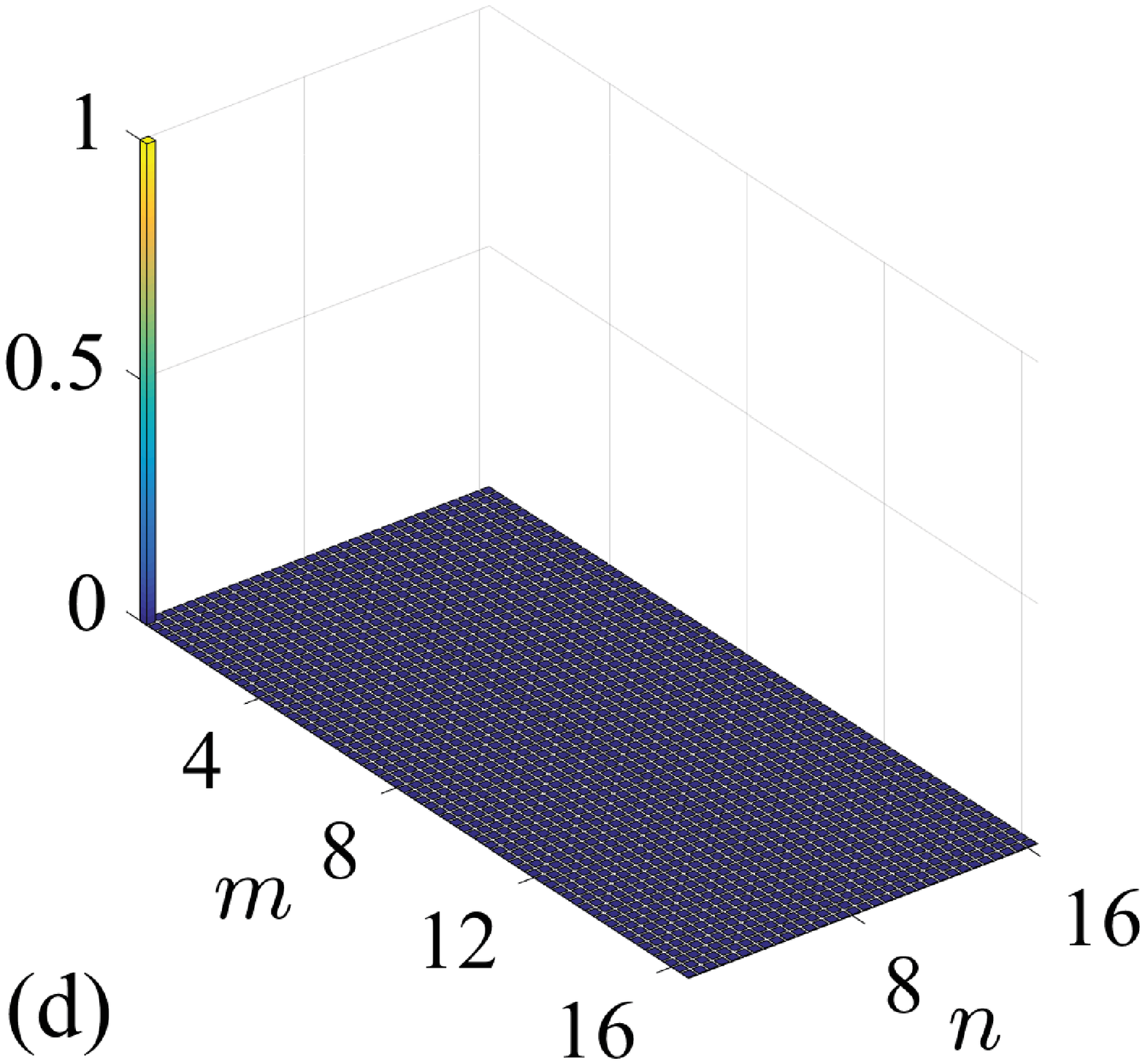} 
\caption{%
	Plots of $\braket{\textsf{v}(m,n,\mu)|E}$
	with intracell label~$\mu$ not explicit
	and
	(a)~$E\approx1.4635$ for~$\chi_{16,16,4}^{\text{r}}$;
	(b)~$E\approx1.4640$ for~$\chi_{16,16,4}^{-}$ for one broken vertex at~$(1,1,1)$;
	(c)~The difference between (a) and (b);
	(d)~The localized eigenstate of~$\chi_{16,16,4}^{-}$ which localize on the broken site.%
	}
\label{fig:brokenplot}
\end{figure}

The reason why a few eigenvalues have shifted can be understood in the context of
Anderson localization~\cite{anderson1958},
which we can see with the help of Fig.~\ref{fig:brokenplot}.
We consider one of eigenstates in the two-dimensional eigenspace corresponding to the eigenvalue $E=1.4635$,
and this eigenstate is depicted in Fig.~\ref{fig:brokenplot}(a).
We observe approximately two cycles at one frequency along the~$m$ axis
and a beat between two frequencies along the~$n$ axis.
Then we isolate, or break, the (1,1,1) vertex and show in Fig.~\ref{fig:brokenplot}(b)
the eigenstate for the lower eigenvalue.
Figures~~\ref{fig:brokenplot}(a,b)
look similar but differ markedly at the broken vertex.
To elucidate this difference,
we plot the arithmetic difference of the two eigenstates in Fig.~\ref{fig:brokenplot}(c),
which shows clearly a spike at the (1,1,1) vertex.
This spike shows that this peak is missing from the eigenstate for the broken-vertex case.

In Fig.~\ref{fig:brokenplot}(d),
we depict the localized state with eigenvalue $E=0$.
This depiction shows the the effect of breaking a vertex,
which is to lead to a completely localized eigenstate at (1,1,1).
This shifting of an eigenvalue and corresponding revision of eigenstates to transition from no support to total support at the defect
(broken vertex)
is a manifestation of Anderson localization.
Thus, the broken-vertex $\chi^-$ graph leads to quantum walks experiencing defects and thus Anderson localization.

\section{Results and discussion}
\label{sec:discussion}
\begin{figure}
  \includegraphics[width=0.35\columnwidth]{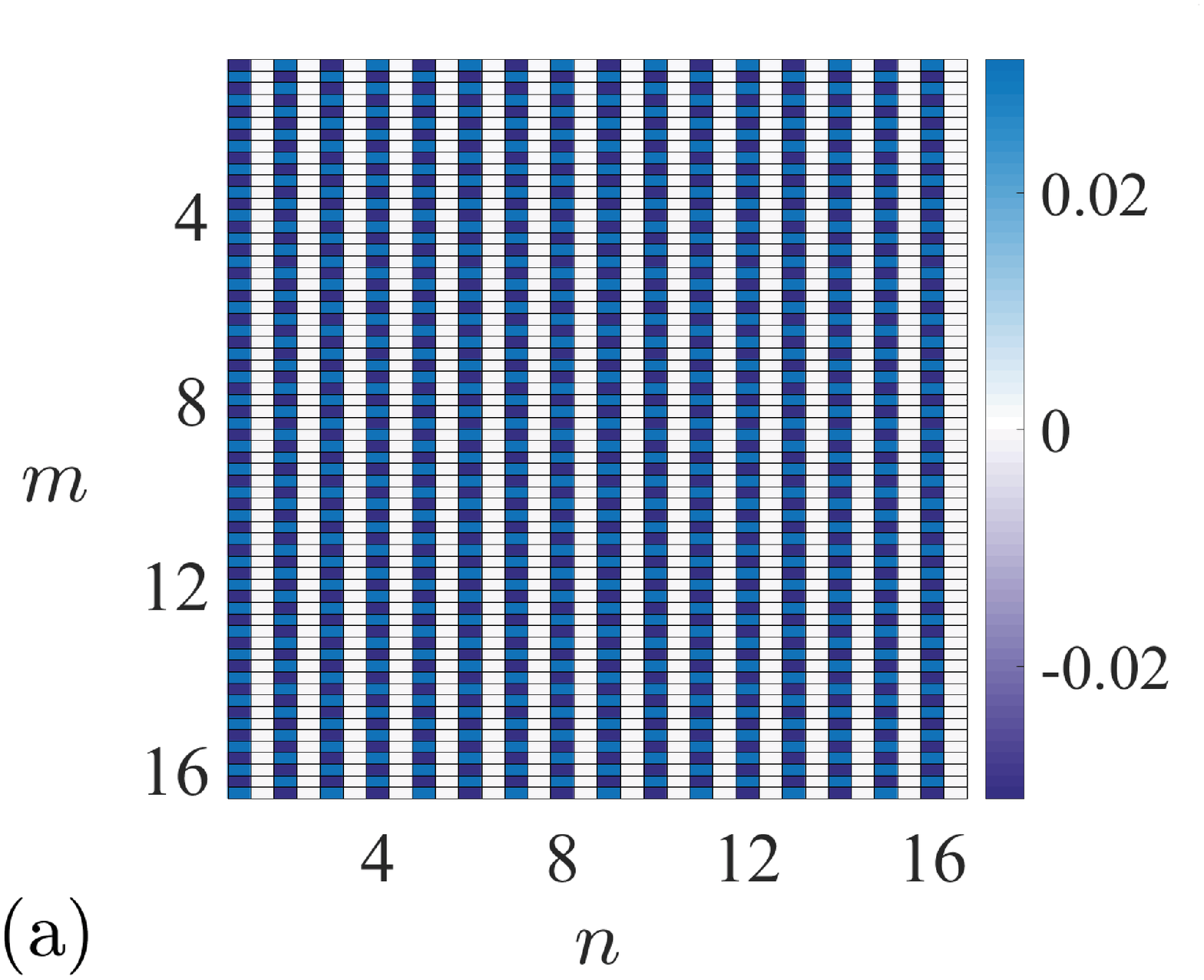}
  \includegraphics[width=0.35\columnwidth]{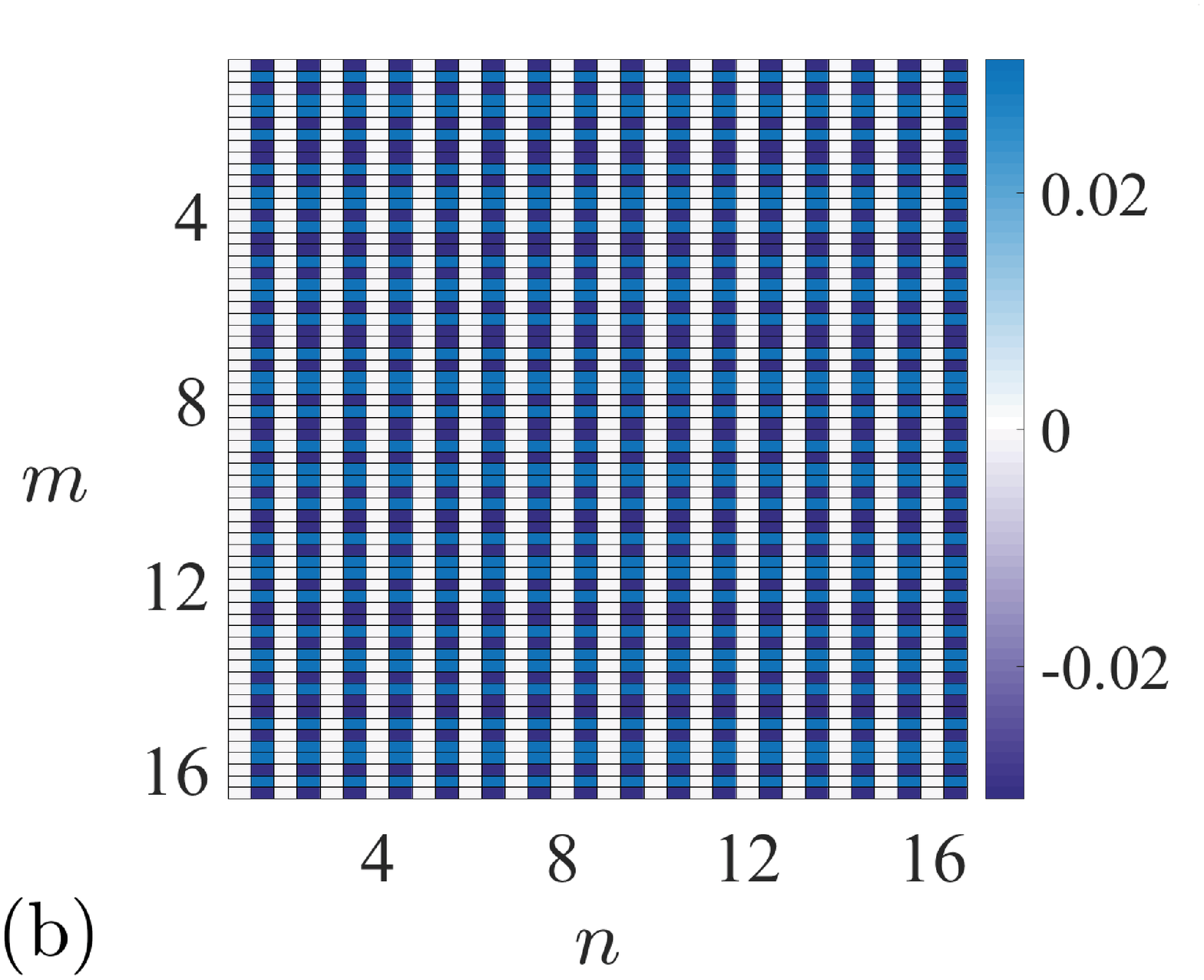}
  \includegraphics[width=0.35\columnwidth]{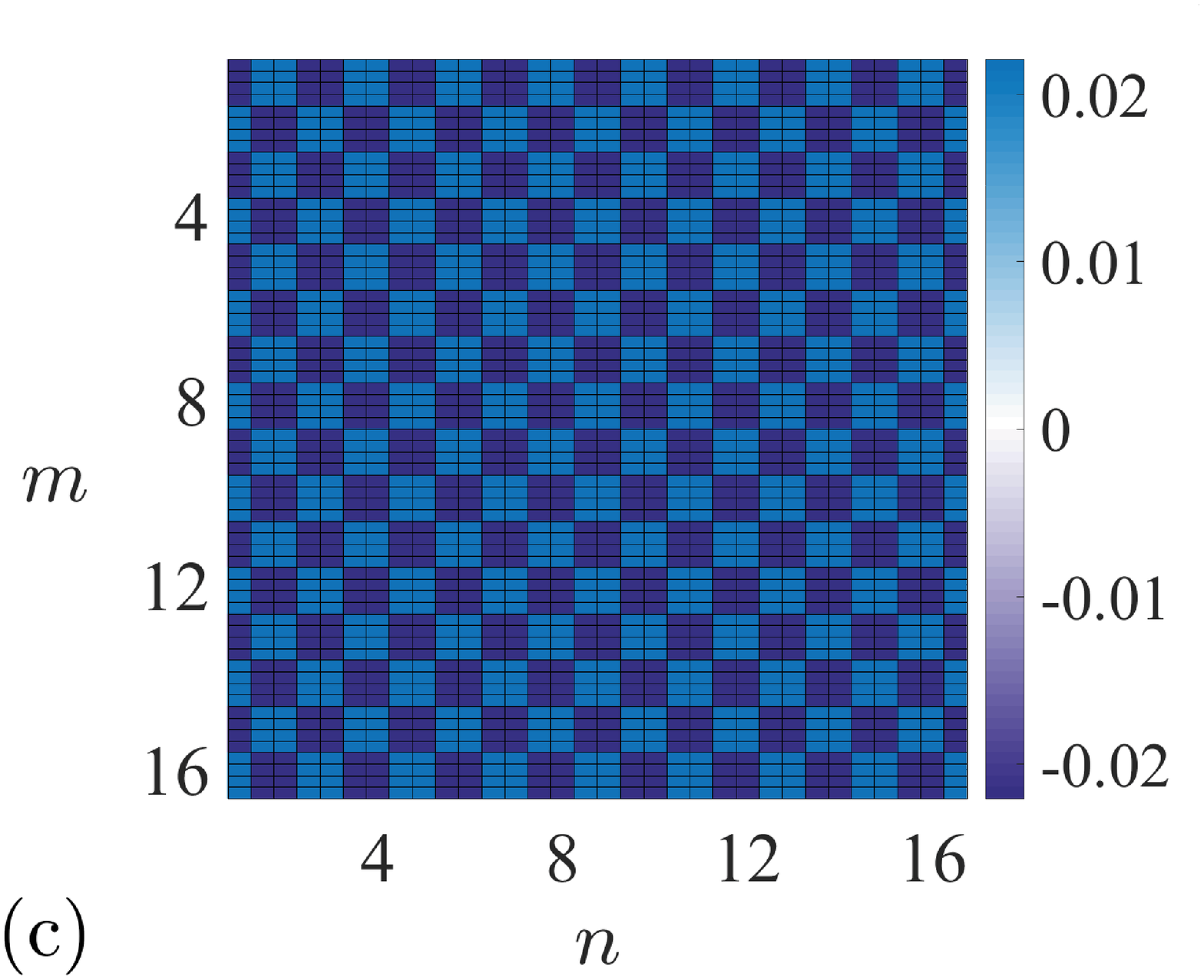}
  \includegraphics[width=0.35\columnwidth]{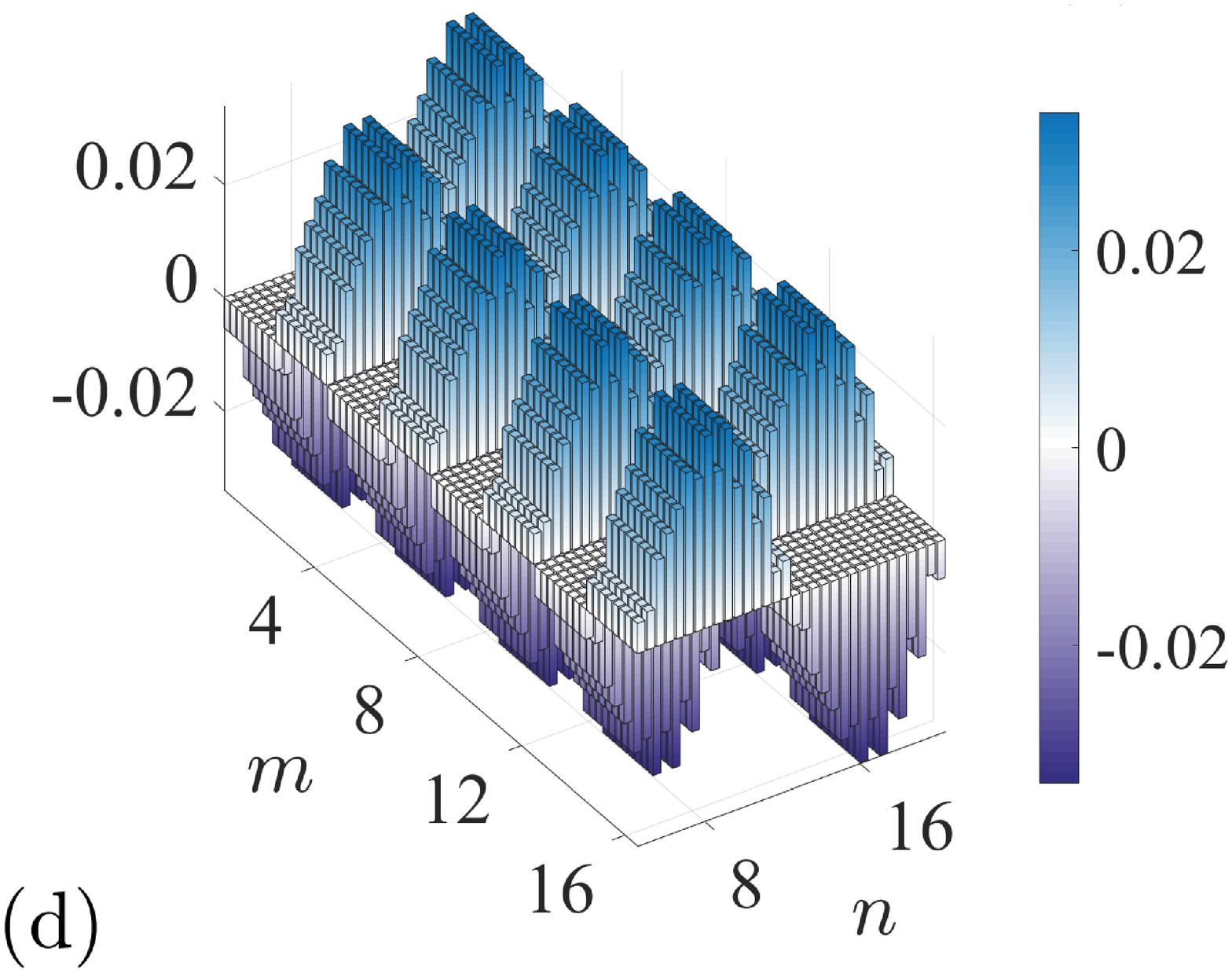}
\caption{%
	   Plots of $\braket{\textsf{v}(m,n,\mu)|E\bm{s}}$
	    for~$\chi_{16,16,4}^{\text{p}}$
	with intracell label~$\mu$ not explicit
	and
	(a)~$\ket{E\bm{s}}_{\triangleright}
		=\ket{4,-1^{\otimes2},1^{\otimes4},-1^{\otimes2}}$,
	(b)~$\ket{E\bm{s}}_{\triangleleft}
		=\ket{4,1^{\otimes2},-1^{\otimes4},1^{\otimes2}}$,
	(c)~$\ket{E\bm{s}}_{\bowtie}
		=\ket{12,1^{\otimes4},-1^{\otimes4}}$ and
	(d)~$\ket{E\bm{s}}_{\bowtie}
		=\ket{6-\zeta_0-\sqrt{16+\zeta_0^2},1^{\otimes4},0,-1,\zeta_0,-1}$
		for $\zeta_0:=\cos(\frac{15}{16}2\pi)$.%
        }
  \label{fig:threeeigen}
\end{figure}

When we lift the degeneracy of the QW~$\hat{H}$,
we analyze salient properties of the eigenstates~$\{\ket{E(\bm{\imath})\bm{s(\bm{\imath})}}\}$ according to the spectra.
First we consider eigenstates~$\{\ket{E(\bm{\imath})\bm{s(\bm{\imath})}}\}$ belonging to the unit lattice~$\mathbb X_{\triangleright}$,
denoted by~$\{\ket{E\bm{s}}_{\triangleright}\}$,
where~$\bm{\imath}\neq 0$ and~$\bm{\imath'}=0$ according to Eq.~(\ref{eq:xleft}).
Through Eqs.~(\ref{eq:s1s2}) and~(\ref{eq:s3s4}),
we have~$s_1\neq 1$ and~$s_3=s_4=1$.
As~$s_3$ and~$s_4$ correspond to intracell right-side translation and mirror permutation symmetry,
if we consider
$\braket{\textsf{v}|E\bm{s}}_{\triangleright}$,
we conclude that
\begin{equation}
\braket{\textsf{v}|E\bm{s}}_{\triangleright}=\braket{\textsf{v}'|E\bm{s}}_{\triangleright},
\end{equation}
for any~$\textsf{v}$ and~$\textsf{v}'$ on the right side of the same unit cell.
Notice that~$s_1$ corresponds to intracell left-side translation symmetry.
If~$s_1\neq 1$,
we must have
\begin{equation}
\braket{\textsf{v}|E\bm{s}}_{\triangleright}=\braket{\textsf{v}'|E\bm{s}}_{\triangleright}=0,
\end{equation}
for any~$\textsf{v}$ and~$\textsf{v}'$ on the right side of the same unit cell.
Through the above analysis, we see that~$\braket{\textsf{v}|E\bm{s}}_{\triangleright}$ has nonzero component only on the left side of the unit cells as shown in Fig.~\ref{fig:threeeigen}(a).
Similarly,
$\braket{\textsf{v}|E\bm{s}}_{\triangleleft}$
has nonzero component only on the right side of the unit cells as shown in Fig.~\ref{fig:threeeigen}(b) and
$\braket{\textsf{v}|E\bm{s}}_{\bowtie}$
has nonzero component on both side of the cells as shown in Fig.~\ref{fig:threeeigen}(c).
Eigenstates are oscillatory in both horizontal and vertical directions
and are effective lattice ``modes.''

We have a comprehensive description of the quantum-walk ``modes'' and thereby analyze the walker's evolution on the~$\chi^\text{p}_{MNL}$ graph,
shown in Fig.~\ref{fig:walkplot}(e) for~$\chi^\text{p}$.
The mode description helps explain features such as the walker's localization in Fig.~\ref{fig:walkplot}.
The walker's state
\begin{equation}
	\ket{\Psi_\textsf{v}(t)}
		=\text{e}^{-\text{i}\hat{H}t}\ket{\textsf{v}}
		=\sum_{\bm{\imath}}\text{e}^{-\text{i}Et}
			\ket{E(\bm{\imath})\bm{s}(\bm{\imath})}
				\braket{E(\bm{\imath})\bm{s}(\bm{\imath})|\textsf{v}}
\end{equation}
is supported by subspaces
$\mathbb X_{\triangleright,\triangleleft,\bowtie}$
yielding orthogonal states
$\ket{\Psi_\textsf{v}(t)}_{\triangleright,\triangleleft,\bowtie}$,
respectively.
Initialization at~$\textsf{v}$ in the left side of a cell yields
\begin{equation}
\label{eq:subspacesl}
	\|\ket{\Psi_\textsf{v}(t)}_{\triangleright}\|^2=\frac{L-1}{L},
	\|\ket{\Psi_\textsf{v}(t)}_{\bowtie}\|^2=\frac{1}{L},
	\|\ket{\Psi_\textsf{v}(t)}_{\triangleleft}\|^2=0.
\end{equation}
Similarly, the walker starting on the right side of the cell
yields
\begin{equation}
\label{eq:subspacesr}
	\|\ket{\Psi_\textsf{v}(t)}_{\triangleleft}\|^2=\frac{L-1}{L},
	\|\ket{\Psi_\textsf{v}(t)}_{\bowtie}\|^2=\frac{1}{L},
	\|\ket{\Psi_\textsf{v}(t)}_{\triangleright}\|^2=0.
\end{equation}
Thus, a walker starting on the left or on the right has vertex distribution supported by at least $\frac{L-1}{L}$;
the remaining distribution is spread over all vertices and thus is a small nonzero ``floor'' for the distribution over all $2MNL$ vertices.

\begin{figure}
  \includegraphics[width=0.45\columnwidth]{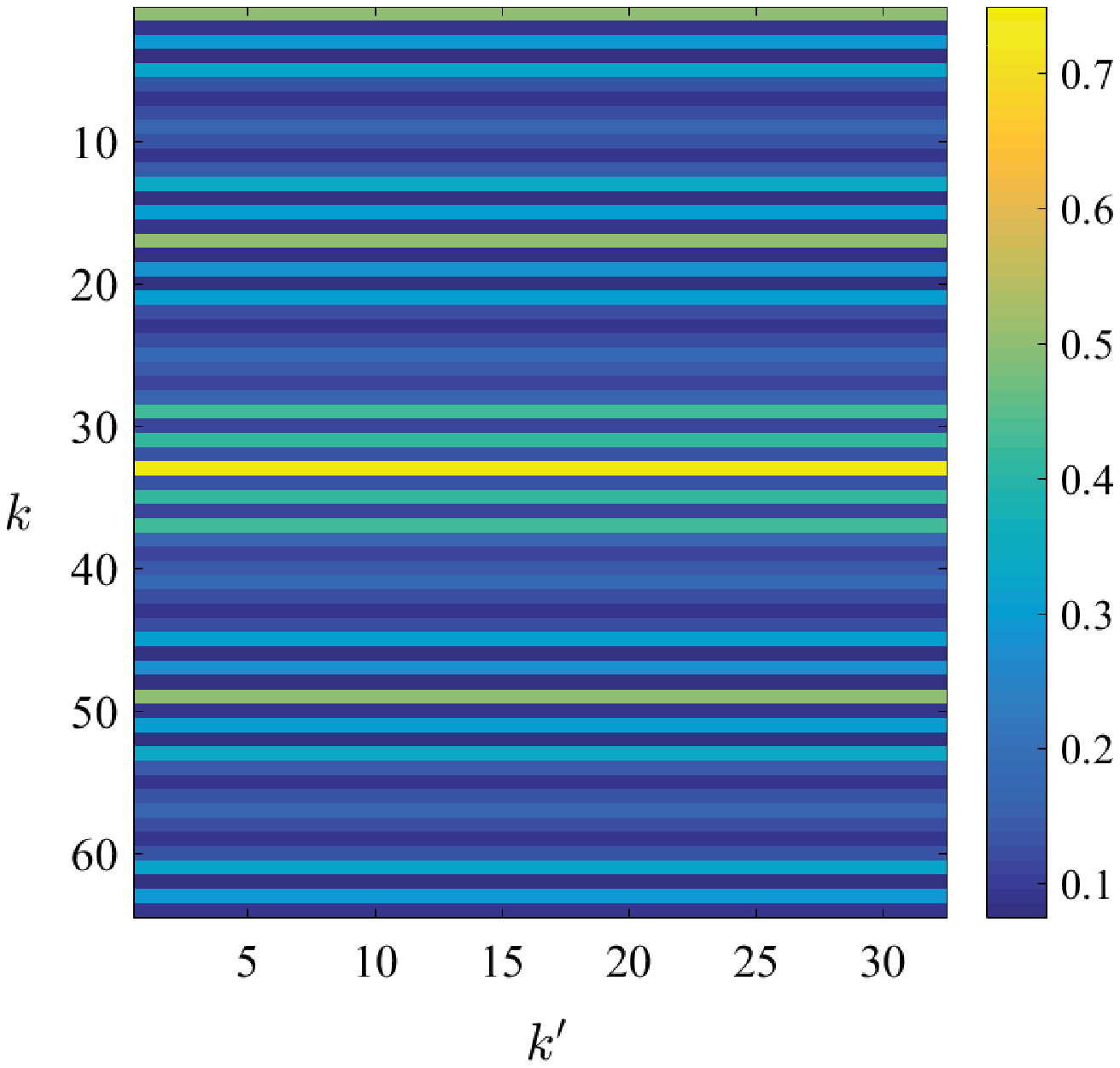}
\caption{%
         Two-dimensional Fourier transform
         	$\|\braket{k,k'|\Psi_\textsf{v}(t=12)}_{\triangleright}\|^{2}$.%
        }
\label{fig:fourier}
\end{figure}

In Fig.~\ref{fig:walkplot}(e) we see that a walker starting at a vertex in the left side of a cell is not only confined to the left side but also stays in just one column with high probability.
This transport is also clear from the two-dimensional Fourier transform
$\|\braket{k,k'|\Psi_\textsf{v}(t)}_{\triangleright}\|^{2}$ in Fig.~\ref{fig:fourier}.
Localization is due to the flat momentum spectrum along the~$k'$ axis,
and the $k$ spectrum is consistent with the momentum distribution for an ordinary QW on a circle~\cite{ambainis2004}.
The walker is initialized on the left of~$\chi_{MNL}^{\text{p}}$
thus has broadband support over all states drawn from
$\mathbb X_{\triangleright}$ resulting in the observed localization to
one column of~$\chi_{MNL}^{\text{p}}$.
This localized walk exhibits
some (indiscernible) leakage outside the column due to $\frac{1}{L}$
support over~$\mathbb X_{\bowtie}$ as seen in Fig.~\ref{fig:walkplot}(e).

The case of the walker localized at a vertex on the right is similar to the case of the walker localized to a cell on the left in that all states parametrized in
$\mathbb X_{\triangleleft}$
are localized to the right side of the initial column cell instead of to the left side of the initial column cell.
A walker commencing at a vertex on the right side has support over
$\mathbb X_{\triangleleft}$
with weight $\frac{L-1}{L}$
and support over $\mathbb X_{\bowtie}$
with weight~$\frac{1}{L}$.
One contrast between localization on the right vs left is evident by comparing Figs.~\ref{fig:walkplot}(a,b).
If all the inter-cell coupling is increased equally,
then the QW only changes by having the walker move more quickly and all the vertical and horizontal localization effects do not change.
That only the speed, and not the features, changes
is easily understood by recognizing that rescaling the coupling strengths does not change the graph symmetries.

The QW can be understood from spectral and stationary-state properties;
now we use this knowledge to examine ~$\chi^+$.
The walker's evolutions on~$\chi^{+}$ and~$\chi^{\text{r}}$ are similar
at the cellular scale,
which is evident by comparing the limiting probability distributions
for~$\chi_{16,16,4}^{+}$ in and~$\chi_{16,16,4}^{\text{r}}$
in Figs.~\ref{fig:walkplot}(d,f), respectively.
This similarity between reflecting and enhanced graphs is readily understood from
Eqs.~(\ref{eq:eigenvalueEr}) and~(\ref{eq:eigenvalueE+}) where we see that
$\chi_{16,16,4}^{+}$ and~$\chi_{16,16,4}^{\text{r}}$ share the same complete set of commuting observables except that energy differs.
In fact,~$\chi_{16,16,4}^{\text{r}}$ has many triply degenerate eigenvalues,
and adding intracell edges to obtain the enhanced graph
causes these triple degeneracies to split into single and double degeneracies that are still concentrated in the same cell.
Thus, the effect of enhancing~$\chi$ by adding intracell edges only spreads the walker within the cell but does not change the QW behaviour on the cellular scale.

Our approach could help with quantum annealer design by making graph symmetries,
and consequences on localization and rate of spreading,
quite clear and explicit.
For example, according to Eqs.~(\ref{eq:subspacesl}) and~(\ref{eq:subspacesr}),
increasing~$L$ causes the walker to become more localized.
Through symmetry analysis we can decide whether to increase or decrease the connections of~$\chi$
in order to lift the degeneracy of the QW~$\hat{H}$
but without significantly changing the eigenstates of the QW~$\hat{H}$.

Although the eigenstate so far are used to analyze properties of the QW,
they can also be prepared.
Preparation of these intricate eigenstates can be achieved by adiabatic evolution~\cite{farhi2000}
beginning with an easy-to-prepare state and then varying~$h$, $j$
and~$k$ in Eq.~(\ref{eq:walkhamiltonian}) as time-dependent labels.
For distinct~$\{a_{\textsf{v}}\}$
the initial Hamiltonian is $\hat{H}(0)=\sum_{\textsf{v}}a_{\textsf{v}}\ket{\textsf{v}}\bra{\textsf{v}}$, and the final Hamiltonian at time~$T$ is
\begin{equation}
	\hat{H}'(T)
		=\hat{H}
			+\sum_{\ell=1}^4\left(y^\ell\hat{S}_{2\ell-1}
				+z^\ell\hat{S}_{2\ell}\right)
\end{equation}
for $y$ a transcendental number and~$z=2$ as an example.
Each initial~$\ket{\textsf{v}}$
evolves to a different mode adiabatically so each mode can be prepared.

\section{Conclusion}
\label{sec:conclusion}
We have studied the QW on the chimera~$\chi$ graph and its variants,
discovered intricate features and showed how a model, or eigenstate,
analysis explains the QW behaviour.
The features of QW localization are explained by spectral analysis,
which builds on completely characterizing the graph symmetries.
We show how these symmetry operators can be incorporated into a target Hamiltonian for adiabatically creating modal states from initial states localized at vertices.

Our analysis of the $\chi$ graph via QWs could possibly aid graph design for quantum annealing as the~$\chi$ graph is used for designing the D-Wave chip.
We caution, though,
that our analysis is based on a single-particle QW,
and quantum annealing is based on exploiting a many-body ground state~\cite{johnson2011}.
Our method thus helps to explore the symmetries of the quantum Hamiltonian with a graph structure that is founded on~$\chi$ or one of its variants
but not to ascertain the many-body ground state or the hardness therein.

Our enhanced~$\chi$ shows that the walker spreads significantly only within the cell and not outside, and our diminished~$\chi$ shows a high probability for a low-error graph to significantly diminish QW dynamics.
Furthermore we have introduced powerful techniques that are useful for studying QWs on other graphs with intrinsic symmetries.
Our approach provides a simple and general way to analyze the quantum walk on complicated
graphs, which have applications to quantum transport
and quantum algorithms, and to designing quantum annealers, which is one of the most significant directions in
practical quantum computing.
An experimental implementation of a two-dimensional quantum walk on the chimera graph could be envisaged for a Bose-Einstein condensate in an optical lattice~\cite{wu2016} or with photon interference~\cite{izaac2017}.
Vacuum noise and thermal effects clearly influence the walker's behaviour and need further investigation.
Our focus has been on the closed system as the closed-system analysis is intricate and reveals much about the role of symmetries and localization in quantum-walk behaviour on the chimera and related graphs.

\begin{acknowledgments}
We thank S.~K.\ Goyal for valuable discussions and acknowledge
China's 1000 Talent Plan and NSFC (Grant No.\ 11675164) for support.
NA acknowledges support from CAS PIFI program (Grant No. 2016PT003)
and BCS appreciates support from Alberta Innovates.
\end{acknowledgments}

\bibliography{biblio}

\begin{thebibliography}{53}%
\makeatletter
\providecommand \@ifxundefined [1]{%
 \@ifx{#1\undefined}
}%
\providecommand \@ifnum [1]{%
 \ifnum #1\expandafter \@firstoftwo
 \else \expandafter \@secondoftwo
 \fi
}%
\providecommand \@ifx [1]{%
 \ifx #1\expandafter \@firstoftwo
 \else \expandafter \@secondoftwo
 \fi
}%
\providecommand \natexlab [1]{#1}%
\providecommand \enquote  [1]{``#1''}%
\providecommand \bibnamefont  [1]{#1}%
\providecommand \bibfnamefont [1]{#1}%
\providecommand \citenamefont [1]{#1}%
\providecommand \href@noop [0]{\@secondoftwo}%
\providecommand \href [0]{\begingroup \@sanitize@url \@href}%
\providecommand \@href[1]{\@@startlink{#1}\@@href}%
\providecommand \@@href[1]{\endgroup#1\@@endlink}%
\providecommand \@sanitize@url [0]{\catcode `\\12\catcode `\$12\catcode
  `\&12\catcode `\#12\catcode `\^12\catcode `\_12\catcode `\%12\relax}%
\providecommand \@@startlink[1]{}%
\providecommand \@@endlink[0]{}%
\providecommand \url  [0]{\begingroup\@sanitize@url \@url }%
\providecommand \@url [1]{\endgroup\@href {#1}{\urlprefix }}%
\providecommand \urlprefix  [0]{URL }%
\providecommand \Eprint [0]{\href }%
\providecommand \doibase [0]{http://dx.doi.org/}%
\providecommand \selectlanguage [0]{\@gobble}%
\providecommand \bibinfo  [0]{\@secondoftwo}%
\providecommand \bibfield  [0]{\@secondoftwo}%
\providecommand \translation [1]{[#1]}%
\providecommand \BibitemOpen [0]{}%
\providecommand \bibitemStop [0]{}%
\providecommand \bibitemNoStop [0]{.\EOS\space}%
\providecommand \EOS [0]{\spacefactor3000\relax}%
\providecommand \BibitemShut  [1]{\csname bibitem#1\endcsname}%
\let\auto@bib@innerbib\@empty
\bibitem [{\citenamefont {Boothby}\ \emph {et~al.}(2016)\citenamefont
  {Boothby}, \citenamefont {King},\ and\ \citenamefont {Roy}}]{boothby2016}%
  \BibitemOpen
  \bibfield  {author} {\bibinfo {author} {\bibfnamefont {T.}~\bibnamefont
  {Boothby}}, \bibinfo {author} {\bibfnamefont {A.~D.}\ \bibnamefont {King}}, \
  and\ \bibinfo {author} {\bibfnamefont {A.}~\bibnamefont {Roy}},\ }\href
  {\doibase 10.1007/s11128-015-1150-6} {\bibfield  {journal} {\bibinfo
  {journal} {Quantum Inf. Process.}\ }\textbf {\bibinfo {volume} {15}},\
  \bibinfo {pages} {495} (\bibinfo {year} {2016})}\BibitemShut {NoStop}%
\bibitem [{\citenamefont {Neukart}\ \emph {et~al.}(2017)\citenamefont
  {Neukart}, \citenamefont {Compostella}, \citenamefont {Seidel}, \citenamefont
  {von Dollen}, \citenamefont {Yarkoni},\ and\ \citenamefont
  {Parney}}]{neukart2017}%
  \BibitemOpen
  \bibfield  {author} {\bibinfo {author} {\bibfnamefont {F.}~\bibnamefont
  {Neukart}}, \bibinfo {author} {\bibfnamefont {G.}~\bibnamefont
  {Compostella}}, \bibinfo {author} {\bibfnamefont {C.}~\bibnamefont {Seidel}},
  \bibinfo {author} {\bibfnamefont {D.}~\bibnamefont {von Dollen}}, \bibinfo
  {author} {\bibfnamefont {S.}~\bibnamefont {Yarkoni}}, \ and\ \bibinfo
  {author} {\bibfnamefont {B.}~\bibnamefont {Parney}},\ }\href@noop {}
  {\bibfield  {journal} {\bibinfo  {journal} {arXiv:1708.01625}\ } (\bibinfo
  {year} {2017})}\BibitemShut {NoStop}%
\bibitem [{\citenamefont {Rosenberg}\ \emph {et~al.}(2016)\citenamefont
  {Rosenberg}, \citenamefont {Haghnegahdar}, \citenamefont {Goddard},
  \citenamefont {Carr}, \citenamefont {Wu},\ and\ \citenamefont
  {de~Prado}}]{rosenberg2016}%
  \BibitemOpen
  \bibfield  {author} {\bibinfo {author} {\bibfnamefont {G.}~\bibnamefont
  {Rosenberg}}, \bibinfo {author} {\bibfnamefont {P.}~\bibnamefont
  {Haghnegahdar}}, \bibinfo {author} {\bibfnamefont {P.}~\bibnamefont
  {Goddard}}, \bibinfo {author} {\bibfnamefont {P.}~\bibnamefont {Carr}},
  \bibinfo {author} {\bibfnamefont {K.}~\bibnamefont {Wu}}, \ and\ \bibinfo
  {author} {\bibfnamefont {M.~L.}\ \bibnamefont {de~Prado}},\ }\href@noop {}
  {\bibfield  {journal} {\bibinfo  {journal} {IEEE J. Sel. Top. Signal
  Process.}\ }\textbf {\bibinfo {volume} {10}},\ \bibinfo {pages} {1053}
  (\bibinfo {year} {2016})}\BibitemShut {NoStop}%
\bibitem [{\citenamefont {Nazareth}\ and\ \citenamefont
  {Spaans}(2015)}]{daryl2015}%
  \BibitemOpen
  \bibfield  {author} {\bibinfo {author} {\bibfnamefont {D.~P.}\ \bibnamefont
  {Nazareth}}\ and\ \bibinfo {author} {\bibfnamefont {J.~D.}\ \bibnamefont
  {Spaans}},\ }\href {http://stacks.iop.org/0031-9155/60/i=10/a=4137}
  {\bibfield  {journal} {\bibinfo  {journal} {Phys. Med. Biol.}\ }\textbf
  {\bibinfo {volume} {60}},\ \bibinfo {pages} {4137} (\bibinfo {year}
  {2015})}\BibitemShut {NoStop}%
\bibitem [{\citenamefont {Ushijima-Mwesigwa}\ \emph {et~al.}(2017)\citenamefont
  {Ushijima-Mwesigwa}, \citenamefont {Negre},\ and\ \citenamefont
  {Mniszewski}}]{ushijima2017}%
  \BibitemOpen
  \bibfield  {author} {\bibinfo {author} {\bibfnamefont {H.}~\bibnamefont
  {Ushijima-Mwesigwa}}, \bibinfo {author} {\bibfnamefont {C.~F.~A.}\
  \bibnamefont {Negre}}, \ and\ \bibinfo {author} {\bibfnamefont {S.~M.}\
  \bibnamefont {Mniszewski}},\ }\href@noop {} {\bibfield  {journal} {\bibinfo
  {journal} {arXiv:1705.03082}\ } (\bibinfo {year} {2017})}\BibitemShut
  {NoStop}%
\bibitem [{\citenamefont {O'Malley}\ \emph {et~al.}(2017)\citenamefont
  {O'Malley}, \citenamefont {Vesselinov}, \citenamefont {Alexandrov},\ and\
  \citenamefont {Alexandrov}}]{omalley2017}%
  \BibitemOpen
  \bibfield  {author} {\bibinfo {author} {\bibfnamefont {D.}~\bibnamefont
  {O'Malley}}, \bibinfo {author} {\bibfnamefont {V.~V.}\ \bibnamefont
  {Vesselinov}}, \bibinfo {author} {\bibfnamefont {B.~S.}\ \bibnamefont
  {Alexandrov}}, \ and\ \bibinfo {author} {\bibfnamefont {L.~B.}\ \bibnamefont
  {Alexandrov}},\ }\href@noop {} {\bibfield  {journal} {\bibinfo  {journal}
  {arXiv:1704.01605}\ } (\bibinfo {year} {2017})}\BibitemShut {NoStop}%
\bibitem [{\citenamefont {Potok}\ \emph {et~al.}(2016)\citenamefont {Potok},
  \citenamefont {Schuman}, \citenamefont {Young}, \citenamefont {Patton},
  \citenamefont {Spedalieri}, \citenamefont {Liu}, \citenamefont {Yao},
  \citenamefont {Rose},\ and\ \citenamefont {Chakma}}]{potok2016}%
  \BibitemOpen
  \bibfield  {author} {\bibinfo {author} {\bibfnamefont {T.~E.}\ \bibnamefont
  {Potok}}, \bibinfo {author} {\bibfnamefont {C.~D.}\ \bibnamefont {Schuman}},
  \bibinfo {author} {\bibfnamefont {S.~R.}\ \bibnamefont {Young}}, \bibinfo
  {author} {\bibfnamefont {R.~M.}\ \bibnamefont {Patton}}, \bibinfo {author}
  {\bibfnamefont {F.}~\bibnamefont {Spedalieri}}, \bibinfo {author}
  {\bibfnamefont {J.}~\bibnamefont {Liu}}, \bibinfo {author} {\bibfnamefont
  {K.-T.}\ \bibnamefont {Yao}}, \bibinfo {author} {\bibfnamefont
  {G.}~\bibnamefont {Rose}}, \ and\ \bibinfo {author} {\bibfnamefont
  {G.}~\bibnamefont {Chakma}},\ }in\ \href {\doibase 10.1109/MLHPC.2016.9}
  {\emph {\bibinfo {booktitle} {Proceedings of the Workshop on Machine Learning
  in High Performance Computing Environments}}}\ (\bibinfo  {publisher} {IEEE
  Press},\ \bibinfo {address} {Piscataway},\ \bibinfo {year} {2016})\ pp.\
  \bibinfo {pages} {47--55}\BibitemShut {NoStop}%
\bibitem [{\citenamefont {Adachi}\ and\ \citenamefont
  {Henderson}(2015)}]{adachi2015}%
  \BibitemOpen
  \bibfield  {author} {\bibinfo {author} {\bibfnamefont {S.~H.}\ \bibnamefont
  {Adachi}}\ and\ \bibinfo {author} {\bibfnamefont {M.~P.}\ \bibnamefont
  {Henderson}},\ }\href@noop {} {\bibfield  {journal} {\bibinfo  {journal}
  {arXiv:1510.06356}\ } (\bibinfo {year} {2015})}\BibitemShut {NoStop}%
\bibitem [{\citenamefont {Choi}(2008)}]{choi2008}%
  \BibitemOpen
  \bibfield  {author} {\bibinfo {author} {\bibfnamefont {V.}~\bibnamefont
  {Choi}},\ }\href {\doibase 10.1007/s11128-008-0082-9} {\bibfield  {journal}
  {\bibinfo  {journal} {Quantum Inf. Process.}\ }\textbf {\bibinfo {volume}
  {7}},\ \bibinfo {pages} {193} (\bibinfo {year} {2008})}\BibitemShut {NoStop}%
\bibitem [{\citenamefont {Choi}(2011)}]{choi2011}%
  \BibitemOpen
  \bibfield  {author} {\bibinfo {author} {\bibfnamefont {V.}~\bibnamefont
  {Choi}},\ }\href {\doibase 10.1007/s11128-010-0200-3} {\bibfield  {journal}
  {\bibinfo  {journal} {Quantum Inf. Process.}\ }\textbf {\bibinfo {volume}
  {10}},\ \bibinfo {pages} {343} (\bibinfo {year} {2011})}\BibitemShut
  {NoStop}%
\bibitem [{\citenamefont {Klymko}\ \emph {et~al.}(2014)\citenamefont {Klymko},
  \citenamefont {Sullivan},\ and\ \citenamefont {Humble}}]{klymko2014}%
  \BibitemOpen
  \bibfield  {author} {\bibinfo {author} {\bibfnamefont {C.}~\bibnamefont
  {Klymko}}, \bibinfo {author} {\bibfnamefont {B.~D.}\ \bibnamefont
  {Sullivan}}, \ and\ \bibinfo {author} {\bibfnamefont {T.~S.}\ \bibnamefont
  {Humble}},\ }\href {\doibase 10.1007/s11128-013-0683-9} {\bibfield  {journal}
  {\bibinfo  {journal} {Quantum Inf. Process.}\ }\textbf {\bibinfo {volume}
  {13}},\ \bibinfo {pages} {709} (\bibinfo {year} {2014})}\BibitemShut
  {NoStop}%
\bibitem [{\citenamefont {Katzgraber}\ \emph {et~al.}(2014)\citenamefont
  {Katzgraber}, \citenamefont {Hamze},\ and\ \citenamefont
  {Andrist}}]{katzgraber2014}%
  \BibitemOpen
  \bibfield  {author} {\bibinfo {author} {\bibfnamefont {H.~G.}\ \bibnamefont
  {Katzgraber}}, \bibinfo {author} {\bibfnamefont {F.}~\bibnamefont {Hamze}}, \
  and\ \bibinfo {author} {\bibfnamefont {R.~S.}\ \bibnamefont {Andrist}},\
  }\href {\doibase 10.1103/PhysRevX.4.021008} {\bibfield  {journal} {\bibinfo
  {journal} {Phys. Rev. X}\ }\textbf {\bibinfo {volume} {4}},\ \bibinfo {pages}
  {021008} (\bibinfo {year} {2014})}\BibitemShut {NoStop}%
\bibitem [{\citenamefont {Vinci}\ \emph {et~al.}(2015)\citenamefont {Vinci},
  \citenamefont {Albash}, \citenamefont {Paz-Silva}, \citenamefont {Hen},\ and\
  \citenamefont {Lidar}}]{vinci2015}%
  \BibitemOpen
  \bibfield  {author} {\bibinfo {author} {\bibfnamefont {W.}~\bibnamefont
  {Vinci}}, \bibinfo {author} {\bibfnamefont {T.}~\bibnamefont {Albash}},
  \bibinfo {author} {\bibfnamefont {G.}~\bibnamefont {Paz-Silva}}, \bibinfo
  {author} {\bibfnamefont {I.}~\bibnamefont {Hen}}, \ and\ \bibinfo {author}
  {\bibfnamefont {D.~A.}\ \bibnamefont {Lidar}},\ }\href {\doibase
  10.1103/PhysRevA.92.042310} {\bibfield  {journal} {\bibinfo  {journal} {Phys.
  Rev. A}\ }\textbf {\bibinfo {volume} {92}},\ \bibinfo {pages} {042310}
  (\bibinfo {year} {2015})}\BibitemShut {NoStop}%
\bibitem [{\citenamefont {Bunyk}\ \emph {et~al.}(2014)\citenamefont {Bunyk},
  \citenamefont {Hoskinson}, \citenamefont {Johnson}, \citenamefont
  {Tolkacheva}, \citenamefont {Altomare}, \citenamefont {Berkley},
  \citenamefont {Harris}, \citenamefont {Hilton}, \citenamefont {Lanting},
  \citenamefont {Przybysz},\ and\ \citenamefont {Whittaker}}]{bunyk2014}%
  \BibitemOpen
  \bibfield  {author} {\bibinfo {author} {\bibfnamefont {P.~I.}\ \bibnamefont
  {Bunyk}}, \bibinfo {author} {\bibfnamefont {E.~M.}\ \bibnamefont
  {Hoskinson}}, \bibinfo {author} {\bibfnamefont {M.~W.}\ \bibnamefont
  {Johnson}}, \bibinfo {author} {\bibfnamefont {E.}~\bibnamefont {Tolkacheva}},
  \bibinfo {author} {\bibfnamefont {F.}~\bibnamefont {Altomare}}, \bibinfo
  {author} {\bibfnamefont {A.~J.}\ \bibnamefont {Berkley}}, \bibinfo {author}
  {\bibfnamefont {R.}~\bibnamefont {Harris}}, \bibinfo {author} {\bibfnamefont
  {J.~P.}\ \bibnamefont {Hilton}}, \bibinfo {author} {\bibfnamefont
  {T.}~\bibnamefont {Lanting}}, \bibinfo {author} {\bibfnamefont {A.~J.}\
  \bibnamefont {Przybysz}}, \ and\ \bibinfo {author} {\bibfnamefont
  {J.}~\bibnamefont {Whittaker}},\ }\href@noop {} {\bibfield  {journal}
  {\bibinfo  {journal} {IEEE Trans. Appl. Supercond.}\ }\textbf {\bibinfo
  {volume} {24}},\ \bibinfo {pages} {1} (\bibinfo {year} {2014})}\BibitemShut
  {NoStop}%
\bibitem [{\citenamefont {Venturelli}\ \emph {et~al.}(2015)\citenamefont
  {Venturelli}, \citenamefont {Mandr\`a}, \citenamefont {Knysh}, \citenamefont
  {O'Gorman}, \citenamefont {Biswas},\ and\ \citenamefont
  {Smelyanskiy}}]{venturelli2015}%
  \BibitemOpen
  \bibfield  {author} {\bibinfo {author} {\bibfnamefont {D.}~\bibnamefont
  {Venturelli}}, \bibinfo {author} {\bibfnamefont {S.}~\bibnamefont
  {Mandr\`a}}, \bibinfo {author} {\bibfnamefont {S.}~\bibnamefont {Knysh}},
  \bibinfo {author} {\bibfnamefont {B.}~\bibnamefont {O'Gorman}}, \bibinfo
  {author} {\bibfnamefont {R.}~\bibnamefont {Biswas}}, \ and\ \bibinfo {author}
  {\bibfnamefont {V.}~\bibnamefont {Smelyanskiy}},\ }\href {\doibase
  10.1103/PhysRevX.5.031040} {\bibfield  {journal} {\bibinfo  {journal} {Phys.
  Rev. X}\ }\textbf {\bibinfo {volume} {5}},\ \bibinfo {pages} {031040}
  (\bibinfo {year} {2015})}\BibitemShut {NoStop}%
\bibitem [{\citenamefont {Aharonov}\ \emph {et~al.}(1993)\citenamefont
  {Aharonov}, \citenamefont {Davidovich},\ and\ \citenamefont
  {Zagury}}]{aharonov1993}%
  \BibitemOpen
  \bibfield  {author} {\bibinfo {author} {\bibfnamefont {Y.}~\bibnamefont
  {Aharonov}}, \bibinfo {author} {\bibfnamefont {L.}~\bibnamefont
  {Davidovich}}, \ and\ \bibinfo {author} {\bibfnamefont {N.}~\bibnamefont
  {Zagury}},\ }\href {\doibase 10.1103/PhysRevA.48.1687} {\bibfield  {journal}
  {\bibinfo  {journal} {Phys. Rev. A}\ }\textbf {\bibinfo {volume} {48}},\
  \bibinfo {pages} {1687} (\bibinfo {year} {1993})}\BibitemShut {NoStop}%
\bibitem [{\citenamefont {Noh}\ and\ \citenamefont {Rieger}(2004)}]{noh2004}%
  \BibitemOpen
  \bibfield  {author} {\bibinfo {author} {\bibfnamefont {J.~D.}\ \bibnamefont
  {Noh}}\ and\ \bibinfo {author} {\bibfnamefont {H.}~\bibnamefont {Rieger}},\
  }\href {\doibase 10.1103/PhysRevLett.92.118701} {\bibfield  {journal}
  {\bibinfo  {journal} {Phys. Rev. Lett.}\ }\textbf {\bibinfo {volume} {92}},\
  \bibinfo {pages} {118701} (\bibinfo {year} {2004})}\BibitemShut {NoStop}%
\bibitem [{\citenamefont {Childs}\ and\ \citenamefont
  {Goldstone}(2004)}]{childs2004}%
  \BibitemOpen
  \bibfield  {author} {\bibinfo {author} {\bibfnamefont {A.~M.}\ \bibnamefont
  {Childs}}\ and\ \bibinfo {author} {\bibfnamefont {J.}~\bibnamefont
  {Goldstone}},\ }\href {\doibase 10.1103/PhysRevA.70.022314} {\bibfield
  {journal} {\bibinfo  {journal} {Phys. Rev. A}\ }\textbf {\bibinfo {volume}
  {70}},\ \bibinfo {pages} {022314} (\bibinfo {year} {2004})}\BibitemShut
  {NoStop}%
\bibitem [{\citenamefont {Childs}(2009)}]{childs2009}%
  \BibitemOpen
  \bibfield  {author} {\bibinfo {author} {\bibfnamefont {A.~M.}\ \bibnamefont
  {Childs}},\ }\href {\doibase 10.1103/PhysRevLett.102.180501} {\bibfield
  {journal} {\bibinfo  {journal} {Phys. Rev. Lett.}\ }\textbf {\bibinfo
  {volume} {102}},\ \bibinfo {pages} {180501} (\bibinfo {year}
  {2009})}\BibitemShut {NoStop}%
\bibitem [{\citenamefont {Lovett}\ \emph {et~al.}(2010)\citenamefont {Lovett},
  \citenamefont {Cooper}, \citenamefont {Everitt}, \citenamefont {Trevers},\
  and\ \citenamefont {Kendon}}]{lovett2010}%
  \BibitemOpen
  \bibfield  {author} {\bibinfo {author} {\bibfnamefont {N.~B.}\ \bibnamefont
  {Lovett}}, \bibinfo {author} {\bibfnamefont {S.}~\bibnamefont {Cooper}},
  \bibinfo {author} {\bibfnamefont {M.}~\bibnamefont {Everitt}}, \bibinfo
  {author} {\bibfnamefont {M.}~\bibnamefont {Trevers}}, \ and\ \bibinfo
  {author} {\bibfnamefont {V.}~\bibnamefont {Kendon}},\ }\href {\doibase
  10.1103/PhysRevA.81.042330} {\bibfield  {journal} {\bibinfo  {journal} {Phys.
  Rev. A}\ }\textbf {\bibinfo {volume} {81}},\ \bibinfo {pages} {042330}
  (\bibinfo {year} {2010})}\BibitemShut {NoStop}%
\bibitem [{\citenamefont {D\"ur}\ \emph {et~al.}(2002)\citenamefont {D\"ur},
  \citenamefont {Raussendorf}, \citenamefont {Kendon},\ and\ \citenamefont
  {Briegel}}]{dur2002}%
  \BibitemOpen
  \bibfield  {author} {\bibinfo {author} {\bibfnamefont {W.}~\bibnamefont
  {D\"ur}}, \bibinfo {author} {\bibfnamefont {R.}~\bibnamefont {Raussendorf}},
  \bibinfo {author} {\bibfnamefont {V.~M.}\ \bibnamefont {Kendon}}, \ and\
  \bibinfo {author} {\bibfnamefont {H.-J.}\ \bibnamefont {Briegel}},\ }\href
  {\doibase 10.1103/PhysRevA.66.052319} {\bibfield  {journal} {\bibinfo
  {journal} {Phys. Rev. A}\ }\textbf {\bibinfo {volume} {66}},\ \bibinfo
  {pages} {052319} (\bibinfo {year} {2002})}\BibitemShut {NoStop}%
\bibitem [{\citenamefont {Z\"ahringer}\ \emph {et~al.}(2010)\citenamefont
  {Z\"ahringer}, \citenamefont {Kirchmair}, \citenamefont {Gerritsma},
  \citenamefont {Solano}, \citenamefont {Blatt},\ and\ \citenamefont
  {Roos}}]{zahringer2010}%
  \BibitemOpen
  \bibfield  {author} {\bibinfo {author} {\bibfnamefont {F.}~\bibnamefont
  {Z\"ahringer}}, \bibinfo {author} {\bibfnamefont {G.}~\bibnamefont
  {Kirchmair}}, \bibinfo {author} {\bibfnamefont {R.}~\bibnamefont
  {Gerritsma}}, \bibinfo {author} {\bibfnamefont {E.}~\bibnamefont {Solano}},
  \bibinfo {author} {\bibfnamefont {R.}~\bibnamefont {Blatt}}, \ and\ \bibinfo
  {author} {\bibfnamefont {C.~F.}\ \bibnamefont {Roos}},\ }\href {\doibase
  10.1103/PhysRevLett.104.100503} {\bibfield  {journal} {\bibinfo  {journal}
  {Phys. Rev. Lett.}\ }\textbf {\bibinfo {volume} {104}},\ \bibinfo {pages}
  {100503} (\bibinfo {year} {2010})}\BibitemShut {NoStop}%
\bibitem [{\citenamefont {Izaac}\ \emph {et~al.}(2017)\citenamefont {Izaac},
  \citenamefont {Zhan}, \citenamefont {Bian}, \citenamefont {Wang},
  \citenamefont {Li}, \citenamefont {Wang},\ and\ \citenamefont
  {Xue}}]{izaac2017}%
  \BibitemOpen
  \bibfield  {author} {\bibinfo {author} {\bibfnamefont {J.~A.}\ \bibnamefont
  {Izaac}}, \bibinfo {author} {\bibfnamefont {X.}~\bibnamefont {Zhan}},
  \bibinfo {author} {\bibfnamefont {Z.}~\bibnamefont {Bian}}, \bibinfo {author}
  {\bibfnamefont {K.}~\bibnamefont {Wang}}, \bibinfo {author} {\bibfnamefont
  {J.}~\bibnamefont {Li}}, \bibinfo {author} {\bibfnamefont {J.~B.}\
  \bibnamefont {Wang}}, \ and\ \bibinfo {author} {\bibfnamefont
  {P.}~\bibnamefont {Xue}},\ }\href {\doibase 10.1103/PhysRevA.95.032318}
  {\bibfield  {journal} {\bibinfo  {journal} {Phys. Rev. A}\ }\textbf {\bibinfo
  {volume} {95}},\ \bibinfo {pages} {032318} (\bibinfo {year}
  {2017})}\BibitemShut {NoStop}%
\bibitem [{\citenamefont {Farhi}\ and\ \citenamefont
  {Gutmann}(1998)}]{farhi1998}%
  \BibitemOpen
  \bibfield  {author} {\bibinfo {author} {\bibfnamefont {E.}~\bibnamefont
  {Farhi}}\ and\ \bibinfo {author} {\bibfnamefont {S.}~\bibnamefont
  {Gutmann}},\ }\href {\doibase 10.1103/PhysRevA.58.915} {\bibfield  {journal}
  {\bibinfo  {journal} {Phys. Rev. A}\ }\textbf {\bibinfo {volume} {58}},\
  \bibinfo {pages} {915} (\bibinfo {year} {1998})}\BibitemShut {NoStop}%
\bibitem [{\citenamefont {Nagaj}(2010)}]{daniel2010}%
  \BibitemOpen
  \bibfield  {author} {\bibinfo {author} {\bibfnamefont {D.}~\bibnamefont
  {Nagaj}},\ }\href {\doibase 10.1063/1.3384661} {\bibfield  {journal}
  {\bibinfo  {journal} {J. Math. Phys.}\ }\textbf {\bibinfo {volume} {51}},\
  \bibinfo {pages} {062201} (\bibinfo {year} {2010})}\BibitemShut {NoStop}%
\bibitem [{\citenamefont {Ambainis}(2004)}]{ambainis2004}%
  \BibitemOpen
  \bibfield  {author} {\bibinfo {author} {\bibfnamefont {A.}~\bibnamefont
  {Ambainis}},\ }\href {\doibase 10.1145/992287.992296} {\bibfield  {journal}
  {\bibinfo  {journal} {SIGACT News}\ }\textbf {\bibinfo {volume} {35}},\
  \bibinfo {pages} {22} (\bibinfo {year} {2004})}\BibitemShut {NoStop}%
\bibitem [{\citenamefont {Mohseni}\ \emph {et~al.}(2008)\citenamefont
  {Mohseni}, \citenamefont {Rebentrost}, \citenamefont {Lloyd},\ and\
  \citenamefont {Aspuru-Guzik}}]{masoud2008}%
  \BibitemOpen
  \bibfield  {author} {\bibinfo {author} {\bibfnamefont {M.}~\bibnamefont
  {Mohseni}}, \bibinfo {author} {\bibfnamefont {P.}~\bibnamefont {Rebentrost}},
  \bibinfo {author} {\bibfnamefont {S.}~\bibnamefont {Lloyd}}, \ and\ \bibinfo
  {author} {\bibfnamefont {A.}~\bibnamefont {Aspuru-Guzik}},\ }\href {\doibase
  10.1063/1.3002335} {\bibfield  {journal} {\bibinfo  {journal} {J. Chem.
  Phys.}\ }\textbf {\bibinfo {volume} {129}},\ \bibinfo {pages} {174106}
  (\bibinfo {year} {2008})}\BibitemShut {NoStop}%
\bibitem [{\citenamefont {Biggerstaff}\ \emph {et~al.}(2016)\citenamefont
  {Biggerstaff}, \citenamefont {Heilmann}, \citenamefont {Zecevik},
  \citenamefont {Gr{\"a}fe}, \citenamefont {Broome}, \citenamefont {Fedrizzi},
  \citenamefont {Nolte}, \citenamefont {Szameit}, \citenamefont {White},\ and\
  \citenamefont {Kassal}}]{biggerstaff2016}%
  \BibitemOpen
  \bibfield  {author} {\bibinfo {author} {\bibfnamefont {D.~N.}\ \bibnamefont
  {Biggerstaff}}, \bibinfo {author} {\bibfnamefont {R.}~\bibnamefont
  {Heilmann}}, \bibinfo {author} {\bibfnamefont {A.~A.}\ \bibnamefont
  {Zecevik}}, \bibinfo {author} {\bibfnamefont {M.}~\bibnamefont {Gr{\"a}fe}},
  \bibinfo {author} {\bibfnamefont {M.~A.}\ \bibnamefont {Broome}}, \bibinfo
  {author} {\bibfnamefont {A.}~\bibnamefont {Fedrizzi}}, \bibinfo {author}
  {\bibfnamefont {S.}~\bibnamefont {Nolte}}, \bibinfo {author} {\bibfnamefont
  {A.}~\bibnamefont {Szameit}}, \bibinfo {author} {\bibfnamefont {A.~G.}\
  \bibnamefont {White}}, \ and\ \bibinfo {author} {\bibfnamefont
  {I.}~\bibnamefont {Kassal}},\ }\href@noop {} {\bibfield  {journal} {\bibinfo
  {journal} {Nat. Commun.}\ }\textbf {\bibinfo {volume} {7}} (\bibinfo {year}
  {2016})}\BibitemShut {NoStop}%
\bibitem [{\citenamefont {Carrega}\ \emph {et~al.}(2016)\citenamefont
  {Carrega}, \citenamefont {Solinas}, \citenamefont {Sassetti},\ and\
  \citenamefont {Weiss}}]{carrega2016}%
  \BibitemOpen
  \bibfield  {author} {\bibinfo {author} {\bibfnamefont {M.}~\bibnamefont
  {Carrega}}, \bibinfo {author} {\bibfnamefont {P.}~\bibnamefont {Solinas}},
  \bibinfo {author} {\bibfnamefont {M.}~\bibnamefont {Sassetti}}, \ and\
  \bibinfo {author} {\bibfnamefont {U.}~\bibnamefont {Weiss}},\ }\href
  {\doibase 10.1103/PhysRevLett.116.240403} {\bibfield  {journal} {\bibinfo
  {journal} {Phys. Rev. Lett.}\ }\textbf {\bibinfo {volume} {116}},\ \bibinfo
  {pages} {240403} (\bibinfo {year} {2016})}\BibitemShut {NoStop}%
\bibitem [{\citenamefont {Kitagawa}\ \emph {et~al.}(2010)\citenamefont
  {Kitagawa}, \citenamefont {Rudner}, \citenamefont {Berg},\ and\ \citenamefont
  {Demler}}]{kitagawa2010}%
  \BibitemOpen
  \bibfield  {author} {\bibinfo {author} {\bibfnamefont {T.}~\bibnamefont
  {Kitagawa}}, \bibinfo {author} {\bibfnamefont {M.~S.}\ \bibnamefont
  {Rudner}}, \bibinfo {author} {\bibfnamefont {E.}~\bibnamefont {Berg}}, \ and\
  \bibinfo {author} {\bibfnamefont {E.}~\bibnamefont {Demler}},\ }\href
  {\doibase 10.1103/PhysRevA.82.033429} {\bibfield  {journal} {\bibinfo
  {journal} {Phys. Rev. A}\ }\textbf {\bibinfo {volume} {82}},\ \bibinfo
  {pages} {033429} (\bibinfo {year} {2010})}\BibitemShut {NoStop}%
\bibitem [{\citenamefont {Asb\'oth}(2012)}]{asboth2012}%
  \BibitemOpen
  \bibfield  {author} {\bibinfo {author} {\bibfnamefont {J.~K.}\ \bibnamefont
  {Asb\'oth}},\ }\href {\doibase 10.1103/PhysRevB.86.195414} {\bibfield
  {journal} {\bibinfo  {journal} {Phys. Rev. B}\ }\textbf {\bibinfo {volume}
  {86}},\ \bibinfo {pages} {195414} (\bibinfo {year} {2012})}\BibitemShut
  {NoStop}%
\bibitem [{\citenamefont {Kendon}(2006)}]{kendon2006}%
  \BibitemOpen
  \bibfield  {author} {\bibinfo {author} {\bibfnamefont {V.~M.}\ \bibnamefont
  {Kendon}},\ }\href {\doibase 10.1098/rsta.2006.1901} {\bibfield  {journal}
  {\bibinfo  {journal} {Phil. Trans. R. Soc. A}\ }\textbf {\bibinfo {volume}
  {364}},\ \bibinfo {pages} {3407} (\bibinfo {year} {2006})}\BibitemShut
  {NoStop}%
\bibitem [{\citenamefont {Santha}(2008)}]{santha2008}%
  \BibitemOpen
  \bibfield  {author} {\bibinfo {author} {\bibfnamefont {M.}~\bibnamefont
  {Santha}},\ }in\ \href {\doibase 10.1007/978-3-540-79228-4_3} {\emph
  {\bibinfo {booktitle} {Proceedings of the Theory and Applications of Models
  of Computation}}}\ (\bibinfo  {publisher} {Springer},\ \bibinfo {address}
  {Berlin},\ \bibinfo {year} {2008})\ pp.\ \bibinfo {pages}
  {31--46}\BibitemShut {NoStop}%
\bibitem [{\citenamefont {M\"ulken}\ and\ \citenamefont
  {Blumen}(2005)}]{mulken2005}%
  \BibitemOpen
  \bibfield  {author} {\bibinfo {author} {\bibfnamefont {O.}~\bibnamefont
  {M\"ulken}}\ and\ \bibinfo {author} {\bibfnamefont {A.}~\bibnamefont
  {Blumen}},\ }\href {\doibase 10.1103/PhysRevE.71.016101} {\bibfield
  {journal} {\bibinfo  {journal} {Phys. Rev. E}\ }\textbf {\bibinfo {volume}
  {71}},\ \bibinfo {pages} {016101} (\bibinfo {year} {2005})}\BibitemShut
  {NoStop}%
\bibitem [{\citenamefont {Bougroura}\ \emph {et~al.}(2016)\citenamefont
  {Bougroura}, \citenamefont {Aissaoui}, \citenamefont {Chancellor},\ and\
  \citenamefont {Kendon}}]{bougroura2016}%
  \BibitemOpen
  \bibfield  {author} {\bibinfo {author} {\bibfnamefont {H.}~\bibnamefont
  {Bougroura}}, \bibinfo {author} {\bibfnamefont {H.}~\bibnamefont {Aissaoui}},
  \bibinfo {author} {\bibfnamefont {N.}~\bibnamefont {Chancellor}}, \ and\
  \bibinfo {author} {\bibfnamefont {V.}~\bibnamefont {Kendon}},\ }\href
  {\doibase 10.1103/PhysRevA.94.062331} {\bibfield  {journal} {\bibinfo
  {journal} {Phys. Rev. A}\ }\textbf {\bibinfo {volume} {94}},\ \bibinfo
  {pages} {062331} (\bibinfo {year} {2016})}\BibitemShut {NoStop}%
\bibitem [{\citenamefont {Childs}\ \emph {et~al.}(2002)\citenamefont {Childs},
  \citenamefont {Farhi},\ and\ \citenamefont {Gutmann}}]{childs2002}%
  \BibitemOpen
  \bibfield  {author} {\bibinfo {author} {\bibfnamefont {A.~M.}\ \bibnamefont
  {Childs}}, \bibinfo {author} {\bibfnamefont {E.}~\bibnamefont {Farhi}}, \
  and\ \bibinfo {author} {\bibfnamefont {S.}~\bibnamefont {Gutmann}},\ }\href
  {\doibase 10.1023/A:1019609420309} {\bibfield  {journal} {\bibinfo  {journal}
  {Quantum Inf. Process.}\ }\textbf {\bibinfo {volume} {1}},\ \bibinfo {pages}
  {35} (\bibinfo {year} {2002})}\BibitemShut {NoStop}%
\bibitem [{\citenamefont {Aharonov}\ \emph {et~al.}(2001)\citenamefont
  {Aharonov}, \citenamefont {Ambainis}, \citenamefont {Kempe},\ and\
  \citenamefont {Vazirani}}]{aharonov2001}%
  \BibitemOpen
  \bibfield  {author} {\bibinfo {author} {\bibfnamefont {D.}~\bibnamefont
  {Aharonov}}, \bibinfo {author} {\bibfnamefont {A.}~\bibnamefont {Ambainis}},
  \bibinfo {author} {\bibfnamefont {J.}~\bibnamefont {Kempe}}, \ and\ \bibinfo
  {author} {\bibfnamefont {U.}~\bibnamefont {Vazirani}},\ }in\ \href {\doibase
  10.1145/380752.380758} {\emph {\bibinfo {booktitle} {Proceedings of the
  Thirty-third Annual ACM Symposium on Theory of Computing}}}\ (\bibinfo
  {publisher} {ACM},\ \bibinfo {address} {New York},\ \bibinfo {year} {2001})\
  pp.\ \bibinfo {pages} {50--59}\BibitemShut {NoStop}%
\bibitem [{\citenamefont {Shikano}\ and\ \citenamefont
  {Katsura}(2010)}]{shikano2010}%
  \BibitemOpen
  \bibfield  {author} {\bibinfo {author} {\bibfnamefont {Y.}~\bibnamefont
  {Shikano}}\ and\ \bibinfo {author} {\bibfnamefont {H.}~\bibnamefont
  {Katsura}},\ }\href {\doibase 10.1103/PhysRevE.82.031122} {\bibfield
  {journal} {\bibinfo  {journal} {Phys. Rev. E}\ }\textbf {\bibinfo {volume}
  {82}},\ \bibinfo {pages} {031122} (\bibinfo {year} {2010})}\BibitemShut
  {NoStop}%
\bibitem [{\citenamefont {M{\"u}lken}\ and\ \citenamefont
  {Blumen}(2011)}]{mulken2011}%
  \BibitemOpen
  \bibfield  {author} {\bibinfo {author} {\bibfnamefont {O.}~\bibnamefont
  {M{\"u}lken}}\ and\ \bibinfo {author} {\bibfnamefont {A.}~\bibnamefont
  {Blumen}},\ }\href {\doibase http://dx.doi.org/10.1016/j.physrep.2011.01.002}
  {\bibfield  {journal} {\bibinfo  {journal} {Phys. Rep.}\ }\textbf {\bibinfo
  {volume} {502}},\ \bibinfo {pages} {37} (\bibinfo {year} {2011})}\BibitemShut
  {NoStop}%
\bibitem [{\citenamefont {Venegas-Andraca}(2012)}]{venegas-andraca2012}%
  \BibitemOpen
  \bibfield  {author} {\bibinfo {author} {\bibfnamefont {S.~E.}\ \bibnamefont
  {Venegas-Andraca}},\ }\href {\doibase 10.1007/s11128-012-0432-5} {\bibfield
  {journal} {\bibinfo  {journal} {Quantum Inf. Process.}\ }\textbf {\bibinfo
  {volume} {11}},\ \bibinfo {pages} {1015} (\bibinfo {year}
  {2012})}\BibitemShut {NoStop}%
\bibitem [{\citenamefont {Ceperley}\ and\ \citenamefont
  {Alder}(1980)}]{ceperley1980}%
  \BibitemOpen
  \bibfield  {author} {\bibinfo {author} {\bibfnamefont {D.~M.}\ \bibnamefont
  {Ceperley}}\ and\ \bibinfo {author} {\bibfnamefont {B.~J.}\ \bibnamefont
  {Alder}},\ }\href {\doibase 10.1103/PhysRevLett.45.566} {\bibfield  {journal}
  {\bibinfo  {journal} {Phys. Rev. Lett.}\ }\textbf {\bibinfo {volume} {45}},\
  \bibinfo {pages} {566} (\bibinfo {year} {1980})}\BibitemShut {NoStop}%
\bibitem [{\citenamefont {White}(1992)}]{white1992}%
  \BibitemOpen
  \bibfield  {author} {\bibinfo {author} {\bibfnamefont {S.~R.}\ \bibnamefont
  {White}},\ }\href {\doibase 10.1103/PhysRevLett.69.2863} {\bibfield
  {journal} {\bibinfo  {journal} {Phys. Rev. Lett.}\ }\textbf {\bibinfo
  {volume} {69}},\ \bibinfo {pages} {2863} (\bibinfo {year}
  {1992})}\BibitemShut {NoStop}%
\bibitem [{\citenamefont {Farhi}\ \emph {et~al.}(2000)\citenamefont {Farhi},
  \citenamefont {Goldstone}, \citenamefont {Gutmann},\ and\ \citenamefont
  {Sipser}}]{farhi2000}%
  \BibitemOpen
  \bibfield  {author} {\bibinfo {author} {\bibfnamefont {E.}~\bibnamefont
  {Farhi}}, \bibinfo {author} {\bibfnamefont {J.}~\bibnamefont {Goldstone}},
  \bibinfo {author} {\bibfnamefont {S.}~\bibnamefont {Gutmann}}, \ and\
  \bibinfo {author} {\bibfnamefont {M.}~\bibnamefont {Sipser}},\ }\href@noop {}
  {\bibfield  {journal} {\bibinfo  {journal} {arXiv:quant-ph/0001106}\ }
  (\bibinfo {year} {2000})}\BibitemShut {NoStop}%
\bibitem [{\citenamefont {Marshall}\ \emph {et~al.}(1979)\citenamefont
  {Marshall}, \citenamefont {Olkin},\ and\ \citenamefont
  {Arnold}}]{marshall1979}%
  \BibitemOpen
  \bibfield  {author} {\bibinfo {author} {\bibfnamefont {A.~W.}\ \bibnamefont
  {Marshall}}, \bibinfo {author} {\bibfnamefont {I.}~\bibnamefont {Olkin}}, \
  and\ \bibinfo {author} {\bibfnamefont {B.~C.}\ \bibnamefont {Arnold}},\
  }\href@noop {} {\emph {\bibinfo {title} {Inequalities: Theory of Majorization
  and its Applications}}},\ \bibinfo {series} {Springer Ser. Statist.}, Vol.\
  \bibinfo {volume} {143}\ (\bibinfo  {publisher} {Springer-Verlag},\ \bibinfo
  {address} {New York},\ \bibinfo {year} {1979})\BibitemShut {NoStop}%
\bibitem [{\citenamefont {King}\ \emph {et~al.}(2017)\citenamefont {King},
  \citenamefont {Yarkoni}, \citenamefont {Raymond}, \citenamefont {Ozfidan},
  \citenamefont {King}, \citenamefont {Nevisi}, \citenamefont {Hilton},\ and\
  \citenamefont {McGeoch}}]{king2017}%
  \BibitemOpen
  \bibfield  {author} {\bibinfo {author} {\bibfnamefont {J.}~\bibnamefont
  {King}}, \bibinfo {author} {\bibfnamefont {S.}~\bibnamefont {Yarkoni}},
  \bibinfo {author} {\bibfnamefont {J.}~\bibnamefont {Raymond}}, \bibinfo
  {author} {\bibfnamefont {I.}~\bibnamefont {Ozfidan}}, \bibinfo {author}
  {\bibfnamefont {A.~D.}\ \bibnamefont {King}}, \bibinfo {author}
  {\bibfnamefont {M.~M.}\ \bibnamefont {Nevisi}}, \bibinfo {author}
  {\bibfnamefont {J.~P.}\ \bibnamefont {Hilton}}, \ and\ \bibinfo {author}
  {\bibfnamefont {C.~C.}\ \bibnamefont {McGeoch}},\ }\href@noop {} {\bibfield
  {journal} {\bibinfo  {journal} {arXiv:1701.04579}\ } (\bibinfo {year}
  {2017})}\BibitemShut {NoStop}%
\bibitem [{\citenamefont {Diestel}(2016)}]{dieste2016}%
  \BibitemOpen
  \bibfield  {author} {\bibinfo {author} {\bibfnamefont {R.}~\bibnamefont
  {Diestel}},\ }\href@noop {} {\emph {\bibinfo {title} {Graph Theory}}},\ Grad.
  Texts in Math.\ (\bibinfo  {publisher} {Springer-Verlag},\ \bibinfo {address}
  {Heidelberg},\ \bibinfo {year} {2016})\BibitemShut {NoStop}%
\bibitem [{\citenamefont {Dickson}\ \emph {et~al.}(2013)\citenamefont
  {Dickson}, \citenamefont {Johnson}, \citenamefont {Amin}, \citenamefont
  {Harris}, \citenamefont {Altomare}, \citenamefont {Berkley}, \citenamefont
  {Bunyk}, \citenamefont {Cai}, \citenamefont {Chapple}, \citenamefont
  {Chavez}, \citenamefont {Cioata}, \citenamefont {Cirip}, \citenamefont
  {deBuen}, \citenamefont {Drew-Brook}, \citenamefont {Enderud}, \citenamefont
  {Gildert}, \citenamefont {Hamze}, \citenamefont {Hilton}, \citenamefont
  {Hoskinson}, \citenamefont {Karimi}, \citenamefont {Ladizinsky},
  \citenamefont {Ladizinsky}, \citenamefont {Lanting}, \citenamefont {Mahon},
  \citenamefont {Neufeld}, \citenamefont {Oh}, \citenamefont {Perminov},
  \citenamefont {Petroff}, \citenamefont {Przybysz}, \citenamefont {Rich},
  \citenamefont {Spear}, \citenamefont {Tcaciuc}, \citenamefont {Thom},
  \citenamefont {Tolkacheva}, \citenamefont {Uchaikin}, \citenamefont {Wang},
  \citenamefont {Wilson}, \citenamefont {Merali},\ and\ \citenamefont
  {Rose}}]{dickson2013}%
  \BibitemOpen
  \bibfield  {author} {\bibinfo {author} {\bibfnamefont {N.~G.}\ \bibnamefont
  {Dickson}}, \bibinfo {author} {\bibfnamefont {M.~W.}\ \bibnamefont
  {Johnson}}, \bibinfo {author} {\bibfnamefont {M.~H.}\ \bibnamefont {Amin}},
  \bibinfo {author} {\bibfnamefont {R.}~\bibnamefont {Harris}}, \bibinfo
  {author} {\bibfnamefont {F.}~\bibnamefont {Altomare}}, \bibinfo {author}
  {\bibfnamefont {A.~J.}\ \bibnamefont {Berkley}}, \bibinfo {author}
  {\bibfnamefont {P.}~\bibnamefont {Bunyk}}, \bibinfo {author} {\bibfnamefont
  {J.}~\bibnamefont {Cai}}, \bibinfo {author} {\bibfnamefont {E.~M.}\
  \bibnamefont {Chapple}}, \bibinfo {author} {\bibfnamefont {P.}~\bibnamefont
  {Chavez}}, \bibinfo {author} {\bibfnamefont {F.}~\bibnamefont {Cioata}},
  \bibinfo {author} {\bibfnamefont {T.}~\bibnamefont {Cirip}}, \bibinfo
  {author} {\bibfnamefont {P.}~\bibnamefont {deBuen}}, \bibinfo {author}
  {\bibfnamefont {M.}~\bibnamefont {Drew-Brook}}, \bibinfo {author}
  {\bibfnamefont {C.}~\bibnamefont {Enderud}}, \bibinfo {author} {\bibfnamefont
  {S.}~\bibnamefont {Gildert}}, \bibinfo {author} {\bibfnamefont
  {F.}~\bibnamefont {Hamze}}, \bibinfo {author} {\bibfnamefont {J.~P.}\
  \bibnamefont {Hilton}}, \bibinfo {author} {\bibfnamefont {E.}~\bibnamefont
  {Hoskinson}}, \bibinfo {author} {\bibfnamefont {K.}~\bibnamefont {Karimi}},
  \bibinfo {author} {\bibfnamefont {E.}~\bibnamefont {Ladizinsky}}, \bibinfo
  {author} {\bibfnamefont {N.}~\bibnamefont {Ladizinsky}}, \bibinfo {author}
  {\bibfnamefont {T.}~\bibnamefont {Lanting}}, \bibinfo {author} {\bibfnamefont
  {T.}~\bibnamefont {Mahon}}, \bibinfo {author} {\bibfnamefont
  {R.}~\bibnamefont {Neufeld}}, \bibinfo {author} {\bibfnamefont
  {T.}~\bibnamefont {Oh}}, \bibinfo {author} {\bibfnamefont {I.}~\bibnamefont
  {Perminov}}, \bibinfo {author} {\bibfnamefont {C.}~\bibnamefont {Petroff}},
  \bibinfo {author} {\bibfnamefont {A.}~\bibnamefont {Przybysz}}, \bibinfo
  {author} {\bibfnamefont {C.}~\bibnamefont {Rich}}, \bibinfo {author}
  {\bibfnamefont {P.}~\bibnamefont {Spear}}, \bibinfo {author} {\bibfnamefont
  {A.}~\bibnamefont {Tcaciuc}}, \bibinfo {author} {\bibfnamefont {M.~C.}\
  \bibnamefont {Thom}}, \bibinfo {author} {\bibfnamefont {E.}~\bibnamefont
  {Tolkacheva}}, \bibinfo {author} {\bibfnamefont {S.}~\bibnamefont
  {Uchaikin}}, \bibinfo {author} {\bibfnamefont {J.}~\bibnamefont {Wang}},
  \bibinfo {author} {\bibfnamefont {A.~B.}\ \bibnamefont {Wilson}}, \bibinfo
  {author} {\bibfnamefont {Z.}~\bibnamefont {Merali}}, \ and\ \bibinfo {author}
  {\bibfnamefont {G.}~\bibnamefont {Rose}},\ }\href
  {http://dx.doi.org/10.1038/ncomms2920} {\bibfield  {journal} {\bibinfo
  {journal} {Nat. Commun.}\ }\textbf {\bibinfo {volume} {4}},\ \bibinfo {pages}
  {1903} (\bibinfo {year} {2013})}\BibitemShut {NoStop}%
\bibitem [{\citenamefont {Keating}\ \emph {et~al.}(2007)\citenamefont
  {Keating}, \citenamefont {Linden}, \citenamefont {Matthews},\ and\
  \citenamefont {Winter}}]{keating2007}%
  \BibitemOpen
  \bibfield  {author} {\bibinfo {author} {\bibfnamefont {J.~P.}\ \bibnamefont
  {Keating}}, \bibinfo {author} {\bibfnamefont {N.}~\bibnamefont {Linden}},
  \bibinfo {author} {\bibfnamefont {J.~C.~F.}\ \bibnamefont {Matthews}}, \ and\
  \bibinfo {author} {\bibfnamefont {A.}~\bibnamefont {Winter}},\ }\href
  {\doibase 10.1103/PhysRevA.76.012315} {\bibfield  {journal} {\bibinfo
  {journal} {Phys. Rev. A}\ }\textbf {\bibinfo {volume} {76}},\ \bibinfo
  {pages} {012315} (\bibinfo {year} {2007})}\BibitemShut {NoStop}%
\bibitem [{\citenamefont {Koll{\'a}r}\ \emph {et~al.}(2015)\citenamefont
  {Koll{\'a}r}, \citenamefont {Kiss},\ and\ \citenamefont {Jex}}]{kollar2015}%
  \BibitemOpen
  \bibfield  {author} {\bibinfo {author} {\bibfnamefont {B.}~\bibnamefont
  {Koll{\'a}r}}, \bibinfo {author} {\bibfnamefont {T.}~\bibnamefont {Kiss}}, \
  and\ \bibinfo {author} {\bibfnamefont {I.}~\bibnamefont {Jex}},\ }\href
  {\doibase 10.1103/PhysRevA.91.022308} {\bibfield  {journal} {\bibinfo
  {journal} {Phys. Rev. A}\ }\textbf {\bibinfo {volume} {91}},\ \bibinfo
  {pages} {022308} (\bibinfo {year} {2015})}\BibitemShut {NoStop}%
\bibitem [{\citenamefont {Ambainis}\ \emph {et~al.}(2016)\citenamefont
  {Ambainis}, \citenamefont {Pr\ifmmode~\bar{u}\else \={u}\fi{}sis},
  \citenamefont {Vihrovs},\ and\ \citenamefont {Wong}}]{ambainis2016}%
  \BibitemOpen
  \bibfield  {author} {\bibinfo {author} {\bibfnamefont {A.}~\bibnamefont
  {Ambainis}}, \bibinfo {author} {\bibfnamefont {K.}~\bibnamefont
  {Pr\ifmmode~\bar{u}\else \={u}\fi{}sis}}, \bibinfo {author} {\bibfnamefont
  {J.}~\bibnamefont {Vihrovs}}, \ and\ \bibinfo {author} {\bibfnamefont
  {T.~G.}\ \bibnamefont {Wong}},\ }\href {\doibase 10.1103/PhysRevA.94.062324}
  {\bibfield  {journal} {\bibinfo  {journal} {Phys. Rev. A}\ }\textbf {\bibinfo
  {volume} {94}},\ \bibinfo {pages} {062324} (\bibinfo {year}
  {2016})}\BibitemShut {NoStop}%
\bibitem [{\citenamefont {Anderson}(1958)}]{anderson1958}%
  \BibitemOpen
  \bibfield  {author} {\bibinfo {author} {\bibfnamefont {P.~W.}\ \bibnamefont
  {Anderson}},\ }\href {\doibase 10.1103/PhysRev.109.1492} {\bibfield
  {journal} {\bibinfo  {journal} {Phys. Rev.}\ }\textbf {\bibinfo {volume}
  {109}},\ \bibinfo {pages} {1492} (\bibinfo {year} {1958})}\BibitemShut
  {NoStop}%
\bibitem [{\citenamefont {Johnson}\ \emph {et~al.}(2011)\citenamefont
  {Johnson}, \citenamefont {Amin}, \citenamefont {Gildert}, \citenamefont
  {Lanting}, \citenamefont {Hamze}, \citenamefont {Dickson}, \citenamefont
  {Harris}, \citenamefont {Berkley}, \citenamefont {Johansson}, \citenamefont
  {Bunyk}, \citenamefont {Chapple}, \citenamefont {Enderud}, \citenamefont
  {Hilton}, \citenamefont {Karimi}, \citenamefont {Ladizinsky}, \citenamefont
  {Ladizinsky}, \citenamefont {Oh}, \citenamefont {Perminov}, \citenamefont
  {Rich}, \citenamefont {Thom}, \citenamefont {Tolkacheva}, \citenamefont
  {Truncik}, \citenamefont {Uchaikin}, \citenamefont {Wang}, \citenamefont
  {Wilson},\ and\ \citenamefont {Rose}}]{johnson2011}%
  \BibitemOpen
  \bibfield  {author} {\bibinfo {author} {\bibfnamefont {M.~W.}\ \bibnamefont
  {Johnson}}, \bibinfo {author} {\bibfnamefont {M.~H.~S.}\ \bibnamefont
  {Amin}}, \bibinfo {author} {\bibfnamefont {S.}~\bibnamefont {Gildert}},
  \bibinfo {author} {\bibfnamefont {T.}~\bibnamefont {Lanting}}, \bibinfo
  {author} {\bibfnamefont {F.}~\bibnamefont {Hamze}}, \bibinfo {author}
  {\bibfnamefont {N.}~\bibnamefont {Dickson}}, \bibinfo {author} {\bibfnamefont
  {R.}~\bibnamefont {Harris}}, \bibinfo {author} {\bibfnamefont {A.~J.}\
  \bibnamefont {Berkley}}, \bibinfo {author} {\bibfnamefont {J.}~\bibnamefont
  {Johansson}}, \bibinfo {author} {\bibfnamefont {P.}~\bibnamefont {Bunyk}},
  \bibinfo {author} {\bibfnamefont {E.~M.}\ \bibnamefont {Chapple}}, \bibinfo
  {author} {\bibfnamefont {C.}~\bibnamefont {Enderud}}, \bibinfo {author}
  {\bibfnamefont {J.~P.}\ \bibnamefont {Hilton}}, \bibinfo {author}
  {\bibfnamefont {K.}~\bibnamefont {Karimi}}, \bibinfo {author} {\bibfnamefont
  {E.}~\bibnamefont {Ladizinsky}}, \bibinfo {author} {\bibfnamefont
  {N.}~\bibnamefont {Ladizinsky}}, \bibinfo {author} {\bibfnamefont
  {T.}~\bibnamefont {Oh}}, \bibinfo {author} {\bibfnamefont {I.}~\bibnamefont
  {Perminov}}, \bibinfo {author} {\bibfnamefont {C.}~\bibnamefont {Rich}},
  \bibinfo {author} {\bibfnamefont {M.~C.}\ \bibnamefont {Thom}}, \bibinfo
  {author} {\bibfnamefont {E.}~\bibnamefont {Tolkacheva}}, \bibinfo {author}
  {\bibfnamefont {C.~J.~S.}\ \bibnamefont {Truncik}}, \bibinfo {author}
  {\bibfnamefont {S.}~\bibnamefont {Uchaikin}}, \bibinfo {author}
  {\bibfnamefont {J.}~\bibnamefont {Wang}}, \bibinfo {author} {\bibfnamefont
  {B.}~\bibnamefont {Wilson}}, \ and\ \bibinfo {author} {\bibfnamefont
  {G.}~\bibnamefont {Rose}},\ }\href {http://dx.doi.org/10.1038/nature10012}
  {\bibfield  {journal} {\bibinfo  {journal} {Nature}\ }\textbf {\bibinfo
  {volume} {473}},\ \bibinfo {pages} {194 EP } (\bibinfo {year}
  {2011})}\BibitemShut {NoStop}%
\bibitem [{\citenamefont {Wu}\ \emph {et~al.}(2016)\citenamefont {Wu},
  \citenamefont {Zhang}, \citenamefont {Sun}, \citenamefont {Xu}, \citenamefont
  {Wang}, \citenamefont {Ji}, \citenamefont {Deng}, \citenamefont {Chen},
  \citenamefont {Liu},\ and\ \citenamefont {Pan}}]{wu2016}%
  \BibitemOpen
  \bibfield  {author} {\bibinfo {author} {\bibfnamefont {Z.}~\bibnamefont
  {Wu}}, \bibinfo {author} {\bibfnamefont {L.}~\bibnamefont {Zhang}}, \bibinfo
  {author} {\bibfnamefont {W.}~\bibnamefont {Sun}}, \bibinfo {author}
  {\bibfnamefont {X.-T.}\ \bibnamefont {Xu}}, \bibinfo {author} {\bibfnamefont
  {B.-Z.}\ \bibnamefont {Wang}}, \bibinfo {author} {\bibfnamefont {S.-C.}\
  \bibnamefont {Ji}}, \bibinfo {author} {\bibfnamefont {Y.}~\bibnamefont
  {Deng}}, \bibinfo {author} {\bibfnamefont {S.}~\bibnamefont {Chen}}, \bibinfo
  {author} {\bibfnamefont {X.-J.}\ \bibnamefont {Liu}}, \ and\ \bibinfo
  {author} {\bibfnamefont {J.-W.}\ \bibnamefont {Pan}},\ }\href {\doibase
  10.1126/science.aaf6689} {\bibfield  {journal} {\bibinfo  {journal}
  {Science}\ }\textbf {\bibinfo {volume} {354}},\ \bibinfo {pages} {83}
  (\bibinfo {year} {2016})}\BibitemShut {NoStop}%
\end{thebibliography}%
\end{document}